\documentclass{PoS} \usepackage{amsmath,booktabs,amssymb}
\usepackage[utf8]{inputenc}
\usepackage{mathtools}
\usepackage[numbers,sort&compress]{natbib}
\usepackage{natbibspacing}
\usepackage{lineno,ulem}
\definecolor{darkgreen}{rgb}{0,.6,0}
\DeclareSymbolFont{greek}{OML}{cmr}{m}{n}
\DeclareMathSymbol{\epsilon}{0}{greek}{"0F}

\title{Form factors for semi-leptonic $B$ decays\footnote{This article combines
two contributions: Edwin Lizarazo ``Semi-leptonic form factors for rare B
decays'' and Oliver~Witzel ``$B$ decays with charming final state.''}}

\ShortTitle{Form factors for semi-leptonic $B$ decays}

\author{Jonathan Flynn$^a$, Taku Izubuchi$^{b,c}$, Andreas J\"uttner$^a$, Taichi Kawanai$^d$,\newline
  Christoph Lehner$^b$, Edwin Lizarazo$^{\ddagger\, a}$, Amarjit Soni$^b$, Justus Tobias Tsang$^{a,e}$,\newline
  Oliver Witzel$^{\,\ddagger e}$ (RBC and UKQCD collaborations) \phantom{\speaker{Edwin Lizarazo and Oliver Witzel}}\\
  \llap{$^a$} Physics and Astronomy, University of Southampton, Southampton SO17 1BJ, UK \\
  \llap{$^b$} Physics Department, Brookhaven National Laboratory, Upton, NY 11973, USA \\
  \llap{$^c$} RIKEN BNL Research Center, Brookhaven National Laboratory, Upton, NY 11973, USA \\
  \llap{$^d$} Forschungszentrum J\"ulich, Institute for Advanced Simulation,\newline
              J\"ulich Supercomputing Centre, 52425 J\"ulich, Germany\\
  \llap{$^e$} Higgs Centre for Theoretical Physics, School of Physics and Astronomy,\newline
              The University of Edinburgh, EH9 3FD, UK \\ E-mail:
\email{e.lizarazo@soton.ac.uk, o.witzel@ed.ac.uk}}

\abstract{Semi-leptonic $B$ decays provide promising channels to test the
Standard Model, search for signs of new physics, or determine fundamental
parameters like CKM matrix elements. We present an update on our calculation of
short distance contributions to GIM suppressed rare $B$ decays focusing in
particular on $B_s\to \phi \ell^+ \ell^-$ decays. Furthermore we show first results for our calculation of $B_{(s)}\to D_{(s)}^{(*)}\ell\nu$ semi-leptonic decays involving $b\to c$ transitions.\newline
Our calculations are based on RBC-UKQCD's 2+1 flavor domain-wall fermion and Iwasaki gauge field configurations featuring three lattice spacings in the range  $1.73$ GeV $\le a^{-1} \le 2.77$ GeV and pion masses down to the physical value. We calculate the form factors by simulating $b$-quarks using the relativistic heavy quark action, create light $u/d$ and $s$ quarks with standard domain-wall kernel, and use optimised M\"obius domain-wall fermions for charm quarks.}

\FullConference{34th annual International Symposium on Lattice Field Theory\\
24-30 July 2016\\ University of Southampton, UK}

\begin{document}

\section{Introduction} 
\begin{figure}[b]
  \centering
  \parbox{0.42\textwidth}{
  \begin{picture}(63,27)(-10,24)
     \put(-4,25){\includegraphics[width=50mm]{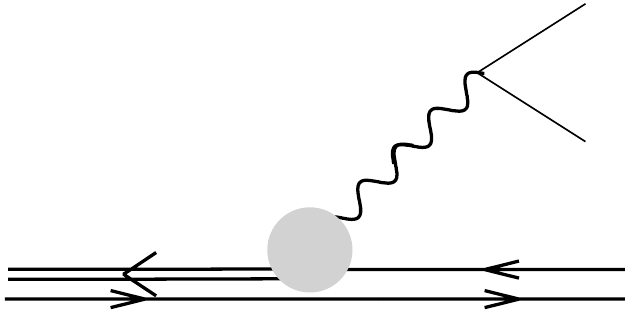}}
     \put(-8,50){a)}
     \put(-10,26){\large{$B$}} \put(47,26){\large{$\pi$}}
     \put(22,37){\large{$W$}}
     \put(43,48){\large{$\ell$}} \put(43.5,37){\large{$\nu_\ell$}}
  \end{picture}}\\[5mm]
\parbox{0.42\textwidth}{
  \begin{picture}(63,27)(52,24)
    \put(60,25){\includegraphics[width=50mm]{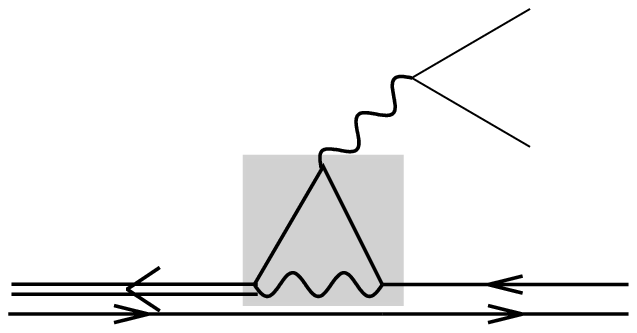}}
    \put(58,49){b)}
    \put(55,26){\large{$B_s$}} \put(111,26){\large{$\phi$}}
    \put(80,33){\large{$t$}} \put(89,33){\large{$t$}}
    \put(82.5,30){\large{$W$}}
    \put(79,42){\large{$Z,\gamma$}}
    \put(102,49){\large{$\ell$}} \put(102,38){\large{$\ell$}}
  \end{picture}}
\hspace{10mm}
\parbox{0.42\textwidth}{
  \begin{picture}(63,28)(55,42)
    \put(60,44){\includegraphics[width=50mm]{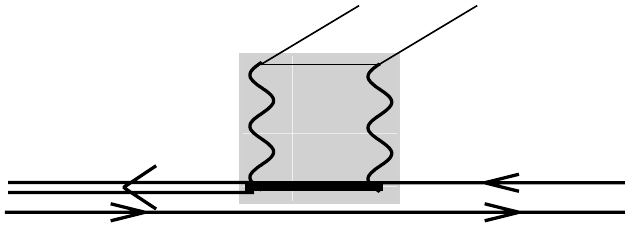}}
    \put(51,67){c)}
    \put(55,45){\large{$B_s$}} \put(110,45){\large{$\phi$}}
    \put(75,51){\large{$W$}} \put(92,51){\large{$W$}}
    \put(89,63){\large{$\ell$}} \put(98.5,63){\large{$\ell$}}
    \put(84,48.5){\large{$t$}} \put(86,57){\large{$\nu$}}
\end{picture}}
\caption{Diagrams sketching the calculation of non-perturbative, short distance
contributions to semi-leptonic decays: a) charged tree-level decay $\big($e.g.,
$B\to\pi \ell \nu$, $B_s\to K\ell\nu$, or $B_{(s)} \to D_{(s)}\ell\nu\big)$, b)
and c) loop-level decays with flavor changing neutral currents $\big($e.g.,
$B_s\to\phi\ell^+\ell^-\big)$. The short distance contributions are indicated
by the gray shading and implemented as point-like operators.}
\label{Fig:DiagramSketches} \end{figure}

Semi-leptonic $B_{(s)}$ decays are receiving considerable attention
both experimentally and theoretically.  They allow to determine 
fundamental parameters of the Standard
Model (SM), like Cabbibo-Kobayashi-Maskawa (CKM) matrix elements providing
 constraints on the SM in searches for new physics. At
tree-level in the SM, only charged flavor changing currents occur and
transitions of bottom quarks to charm or up quarks are suppressed by the small
size of the corresponding CKM matrix elements. Those decays are depicted by
sketch a) in Fig.~\ref{Fig:DiagramSketches}. Transitions of bottom quarks to
down or strange quarks may occur in the SM only at loop level (see sketches b) and c) in Fig.~\ref{Fig:DiagramSketches}) and the corresponding flavor changing neutral currents (FCNC) are further suppressed due to the Glashow-Iliopoulos-Maiani (GIM) mechanism \cite{Glashow:1970gm}.
Since in
the SM these transitions are highly suppressed, they are good candidates to
search for new physics. Anomalies have been reported
comparing SM predictions and experimental results, e.g., for angular observables
\cite{Aaij:2015oid}, branching fractions \cite{Aaij:2015esa}, and the ratio
$R_K$ \cite{Aaij:2014ora}, but are also observed in charged tree-level $b\to c$ transitions see e.g.~\cite{Fajfer:2012vx,Bailey:2012jg,Lees:2012xj,Nandi:2016wlp}.

Besides experimental measurements, theoretical determinations of form factors are needed to test the SM. After carrying out an operator product expansion (OPE), we conventionally classify terms into ``short'' and ``long'' distance contributions. In the following we solely focus on the computation of short distance effects  as one ingredient to better constrain the SM. On the theory side the uncertainties of these short distance contributions arise predominately from hadronic effects.
 Using lattice quantum chromodynamics
(QCD) techniques, we have set-up a program to determine form factors for
semi-leptonic $B_{(s)}$ decays. This program started by calculating semi-leptonic
form factors for $B\to\pi\ell\nu$ and $B_s\to K\ell\nu$ \cite{Flynn:2015mha}
and last year we reported on generalizing our code to additionally
compute GIM suppressed decays with hadronic pseudoscalar or vector final states
\cite{Flynn:2015ynk}. Here we will present updates on our calculation involving
FCNC and, furthermore, present first results for bottom quarks transitioning to
charm quarks. When combined with experimental measurements, $B\to
D^{(*)}\ell\nu$ form factors allow to determine the CKM matrix element
$|V_{cb}|$ but will also allow to compute ratios of branching fractions $R_D^{(*)}$
\begin{align}
  R_{D^{(*)}} = \frac{{\cal B}(D\to D^{(*)}\tau \nu_\tau)}{{\cal B}(D\to D^{(*)}\ell \nu_\ell)},
  \qquad \text{with}\quad \ell=e,\,\mu.
\end{align}
Only very recently the long-standing $2-3 \sigma$ discrepancy
between inclusive and exclusive determinations of $|V_{cb}|$ seems to disappear \cite{Bigi:2016mdz}. The tension between SM predictions and experimental findings for $R_D$
and $R_{D^*}$ \cite{Fajfer:2012vx,Bailey:2012jg,Lees:2012xj,Nandi:2016wlp} are however independent of $|V_{cb}|$ and warrant further investigations. Independent lattice determinations of $B\to D^{(*)}\ell\nu$ semi-leptonic form factors will help to do so.

The remainder of this article is organized as follows: In the next Section we
present the set-up of our computation and give details on the actions and
ensembles used in our simulations. In Section \ref{Sec.RareDecays} we report on
our progress to compute rare semi-leptonic $B$-decays mediated by FCNCs. Due to
the limited space, we will focus here on $B_s\to\phi\ell^+\ell^-$ decays.
In Section \ref{Sec.Charming} we present our setup for $B_{(s)} \to D_{(s)}^{(*)} \ell \nu$
decays with charm quarks discretized as DWF. Finally we give a brief outlook and conclude.

\section{Computational set-up}\label{Sec.Setup}
\begin{table}[t] \centering
  \begin{tabular}{cccccccccc} \toprule
    & &$a^{-1}$ &  &  &   & $M_\pi$  & & \# time \\
    $L^3 \times T$ &$L_s$& [GeV] & $am_l$ & $am_h$ & $am_s^\text{phys}$  & [MeV] & \# configs& sources\\\midrule
    $24^3 \times 64$ &16&1.785(5) & 0.005& 0.040 & 0.03224(18) & 340 &1636 & 1\\
    $24^3 \times 64$ &16&1.785(5) & 0.010 &0.040 & 0.03224(18) & 434 &1419 & 1\\ \midrule
    $32^3 \times 64$ &16&2.383(9) & 0.004 & 0.030 & 0.02477(18) & 302 &628  & 2\\
    $32^3 \times 64$ &16&2.383(9) & 0.006 & 0.030 & 0.02477(18) & 360 &889  & 2\\
    $32^3 \times 64$ &16&2.383(9) & 0.008 & 0.030 & 0.02477(18) & 411 &544  & 2\\ \midrule
    $48^3 \times 96$ &24&1.730(4) & 0.00078 & 0.0362 & 0.03580(16) & 139 &40 & 162\\
    $48^3 \times 96$ &12&2.77(1)\phantom{0} & 0.002144 & 0.02144 & 0.02132(17) & 234 &50 & 24\\
    \bottomrule
  \end{tabular}
  \caption{Overview of the used gauge field ensembles. The ensembles were generated by the RBC and UKQCD collaborations \cite{Allton:2008pn,Aoki:2010dy,Blum:2014tka,CharmPaper} using 2+1 flavor domain-wall fermions and Iwasaki gauge actions. The domain-wall height for light and strange quarks is $M_5=1.8$. The $24^3$ and $32^3$ ensembles are generated using the Shamir domain-wall kernel \cite{Shamir:1993zy,Furman:1994ky}, while $48^3$ ensembles use the M\"obius kernel with $\alpha=2$ \cite{Brower:2012vk}.  Values for the inverse lattice spacing and the quark and meson masses are taken from the refined analysis \cite{Blum:2014tka} and updated to include the finer $a^{-1}=2.77$ GeV ensemble \cite{CharmPaper}. The light sea-quark mass is labeled $am_l$, the heavy sea-quark mass $am_h$, and $am_s^\text{phys}$ is the mass of the physical strange quark mass. The valence strange quark masses used in our simulations on $24^3$ ($32^3$) ensembles are $am_s^\text{sim} = 0.03224\, (0.025)$, while on $48^3$ ensembles we used $am_s^\text{sim} = am_h$. Generation of propagators and calculating contractions is ongoing on both $48^3$ ensembles.}
  \label{tab:1}
\end{table}

Our simulations are based on RBC-UKQCD's set of 2+1 flavor gauge field
configurations \cite{Allton:2008pn, Aoki:2010dy, Blum:2014tka, CharmPaper}
generated with the Iwasaki gauge \cite{Iwasaki:1983ck} and the domain-wall
fermion action \cite{Shamir:1993zy,Furman:1994ky,Brower:2012vk}. Currently our
measurements have been completed on the five ensembles at inverse lattice
spacings of 1.785 and 2.383 GeV featuring unitary pion masses down to $\sim$300
MeV. Work is in progress to include additional data on the new $48^3$ ensemble
with $a^{-1}=2.77$ GeV and the $48^3$ ensemble featuring physical pion masses at $1.730$ GeV. For all ensembles $M_\pi\, L$ is greater than 3.8 and the spatial box sizes are at least 2.6 fm. Details of the used
configurations as well as the number of gauge field configurations and sources per
configuration are summarized in Tab.~\ref{tab:1}. In order to reduce
autocorrelations between lattices, we perform a random 4-vector shift of the
gauge field prior to starting our calculation.

In the valence sector we generate light and strange quark propagators using the
same domain-wall fermion formulation as has been used in the sea-sector. We
simulate the heavy $b$-quarks using the Fermilab \cite{ElKhadra:1996mp} or
relativistic heavy quark (RHQ) action \cite{Lin:2006ur,Christ:2006us}. The RHQ
action is an effective action based on the anisotropic Sheikoleslami-Wohlert
action \cite{Sheikholeslami:1985ij} with a special interpretation of its three
parameters ensuring that discretization errors are small. For this work we
repeated the non-perturbative tuning of the three RHQ parameters following our
prescription published in Ref.~\cite{Aoki:2012xaa}. Re-tuning the RHQ
parameters was triggered by the refined global fit updating the determinations
of lattice spacings \cite{Blum:2014tka} which enables us to consistently
include the newer $48^3$ ensembles. Details of the new RHQ parameters will be
published in a forthcoming paper.

Our choices for calculating 2-point and 3-point correlation functions on the
lattice are guided by our calculation of $B\to\pi\ell\nu$ and $B_s\to K\ell\nu$
semi-leptonic form factors \cite{Flynn:2015mha}. We choose point-sources for
the light and strange quarks, but Gaussian smeared sources \cite{Alford:1995dm} for the heavy bottom
quarks. Since the separation of source and sink is crucial for obtaining a good
signal in the 3-point correlators, we carried out a dedicated study checking
for the signal of all form factors contributing to rare $B$ decays in
\cite{Flynn:2015ynk}. Our previous choices of $t_\text{sink} - t_\text{source}
= 20$ for ensembles with $a^{-1}=1.785$ GeV were confirmed. Scaling that
value proportional to the lattice spacing we use $t_\text{sink} - t_\text{source} = 26 (30)$ for  $a^{-1}=2.383$ (2.77) GeV.

Data presented in the following Sections are analyzed using single elimination
jackknife re-sampling after first averaging correlators computed
with different sources on the same gauge field configuration.

\section{Rare $B$ decays with FCNC}\label{Sec.RareDecays}

The effective Hamiltonian for  weak $b\to q\ell^+\ell^-$ decays (with $q$ a
down or a strange quark and $\ell$ an $e,\,\mu,$ or $\tau$ lepton) is given by \cite{Grinstein:1987vj,Grinstein:1990tj,
Buras:1993xp,Ciuchini:1993ks,Ciuchini:1993fk,Ciuchini:1994xa},
\begin{linenomath*}
  \begin{equation} \mathcal{H}^{b\to q}_\text{eff}  = -\frac{4G_F}{\sqrt{2}}V^*_{tq}V_{tb} \sum_{i=1}^{10} \left(C_i O_i + C^\prime_i O^\prime_i\right),
  \end{equation}
\end{linenomath*}
where  $V_{tq}^*$ and $V_{tb}$ are CKM matrix elements, $O_i^{(\prime)}$ are local operators and $C_i^{(\prime)}$ their corresponding Wilson coefficients determined in \cite{Buras:2002tp,Gambino:2003zm,Altmannshofer:2008dz}. Primed operators differ from unprimed ones by their chirality and are even further suppressed in the SM. Short distance contributions are dominated by dileptonic operators (corresponding to Fig.~\ref{Fig:DiagramSketches}c)

\begin{linenomath*}
  \begin{align}
    O_9 &= \frac{e^2}{16\pi^2} \bar{q}\gamma^\mu P_{L} b \bar{\ell}\gamma_\mu \ell,
    & &O_{10} = \frac{e^2}{16\pi^2} \bar{q}\gamma^\mu P_{L} b \bar{\ell}\gamma_\mu\gamma_5\ell,
    \label{eq:tf-2}
  \end{align}
\end{linenomath*}
and the electromagnetic operator (Fig.~\ref{Fig:DiagramSketches}b)
\begin{linenomath*}
  \begin{equation}
    O_7 = \frac{m_be}{16\pi^2}\bar{q}\sigma^{\mu\nu}P_{R} b F_{\mu\nu}.  \label{eq:tf-1}
  \end{equation}
\end{linenomath*}
In Equations (\ref{eq:tf-2}) and (\ref{eq:tf-1}) the lepton is denoted by $\ell$, the mass
of the $b$-quark by $m_b$ and $P_{L(R)} = \frac{1}{2}(1\mp\gamma^5)$, $\sigma^{\mu\nu}=\frac{i}{2}[\gamma^\mu, \gamma^\nu]$, $F^{\mu\nu} = \partial^\mu A^\nu - \partial^\nu A^\mu$.  Long distance contributions are commonly estimated perturbatively \cite{Grinstein:2004vb, Beylich:2011aq} but their reliability has been questioned due to the presence of charm resonances arising from 4-quark operators also present at loop-level \cite{Lyon:2014hpa}. 
In the following we focus on the computation of the dominant short distance
operators for which a lattice calculation is suitable. We restrict ourselves to
the computation of pseudoscalar $B_{(s)}$ meson decays into a  pseudoscalar or vector meson. Vector mesons are treated as stable
using the narrow width approximation.
To date only the Cambridge group
\cite{Horgan:2013pva,Horgan:2013hoa,Horgan:2015vla} has carried out a lattice
determination for semi-leptonic $B_s\to\phi\ell^+\ell^-$ form factors using
MILC's set of Asqtad gauge field configurations. Further GIM suppressed rare
$B$-decays have been investigated by HPQCD \cite{Bouchard:2013mia}, and 
Fermilab/MILC \cite{Bailey:2015dka,Du:2015tda}. 
%

\begin{figure}[t]
  \centering
  \includegraphics{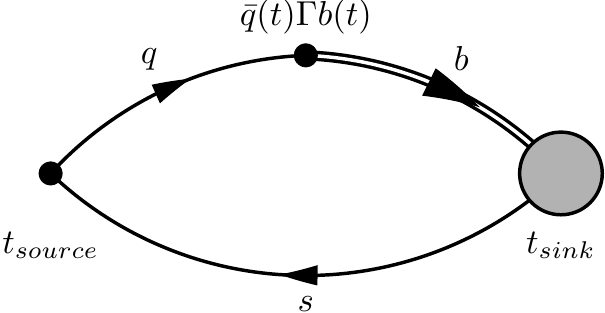}
  \caption{Three point correlator function used to obtain the $B_s\to \phi$ and $B_s\to D_s$ form factors. Double lines correspond to $b$-quark propagators, light, strange and charm propagators are denoted with single lines.}
  \label{fig:3pt}
\end{figure}
\begin{figure}[p]
\centering
\parbox{0.495\textwidth}{\includegraphics[width=\linewidth]{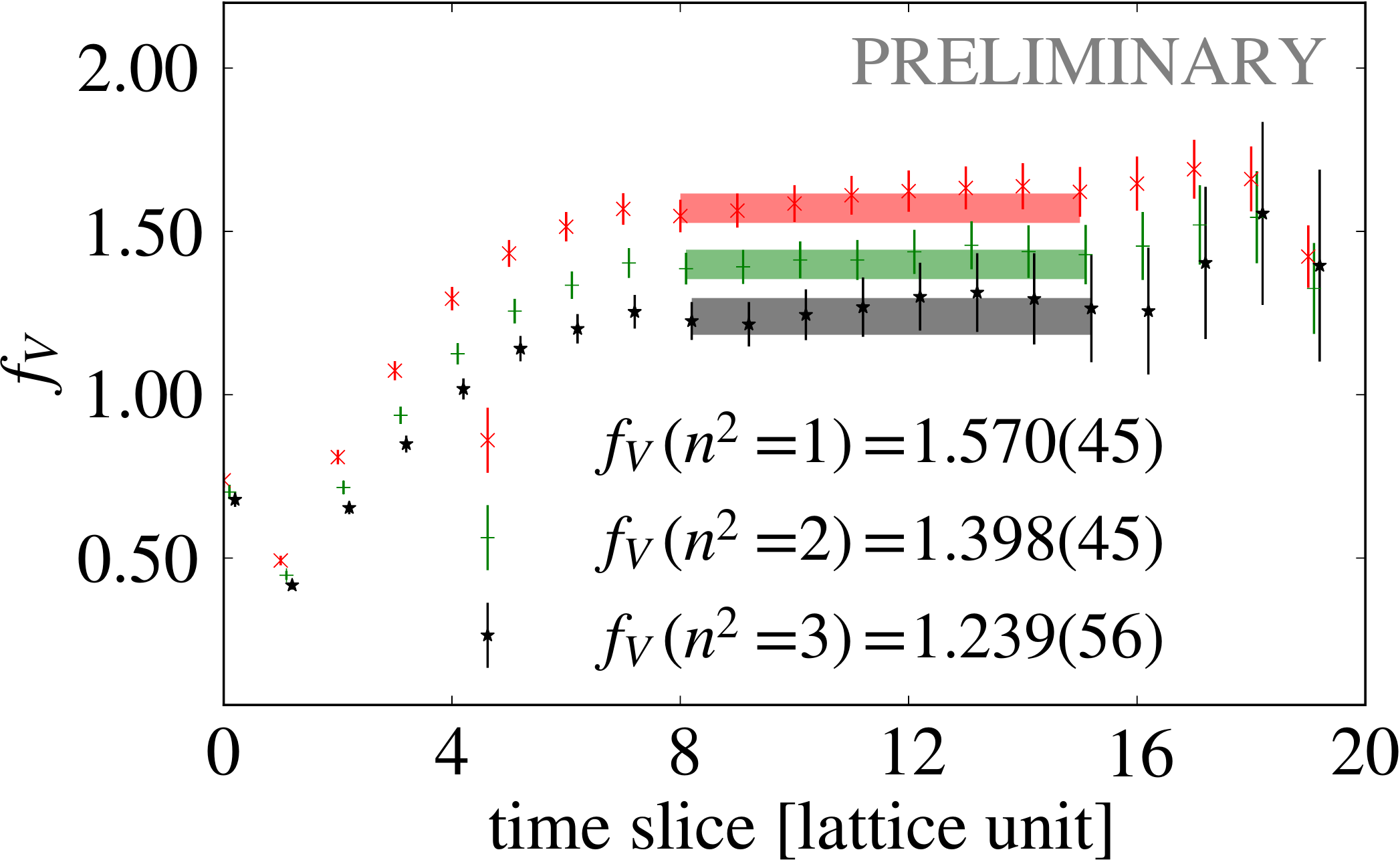}}
\parbox{0.495\textwidth}{\includegraphics[width=\linewidth]{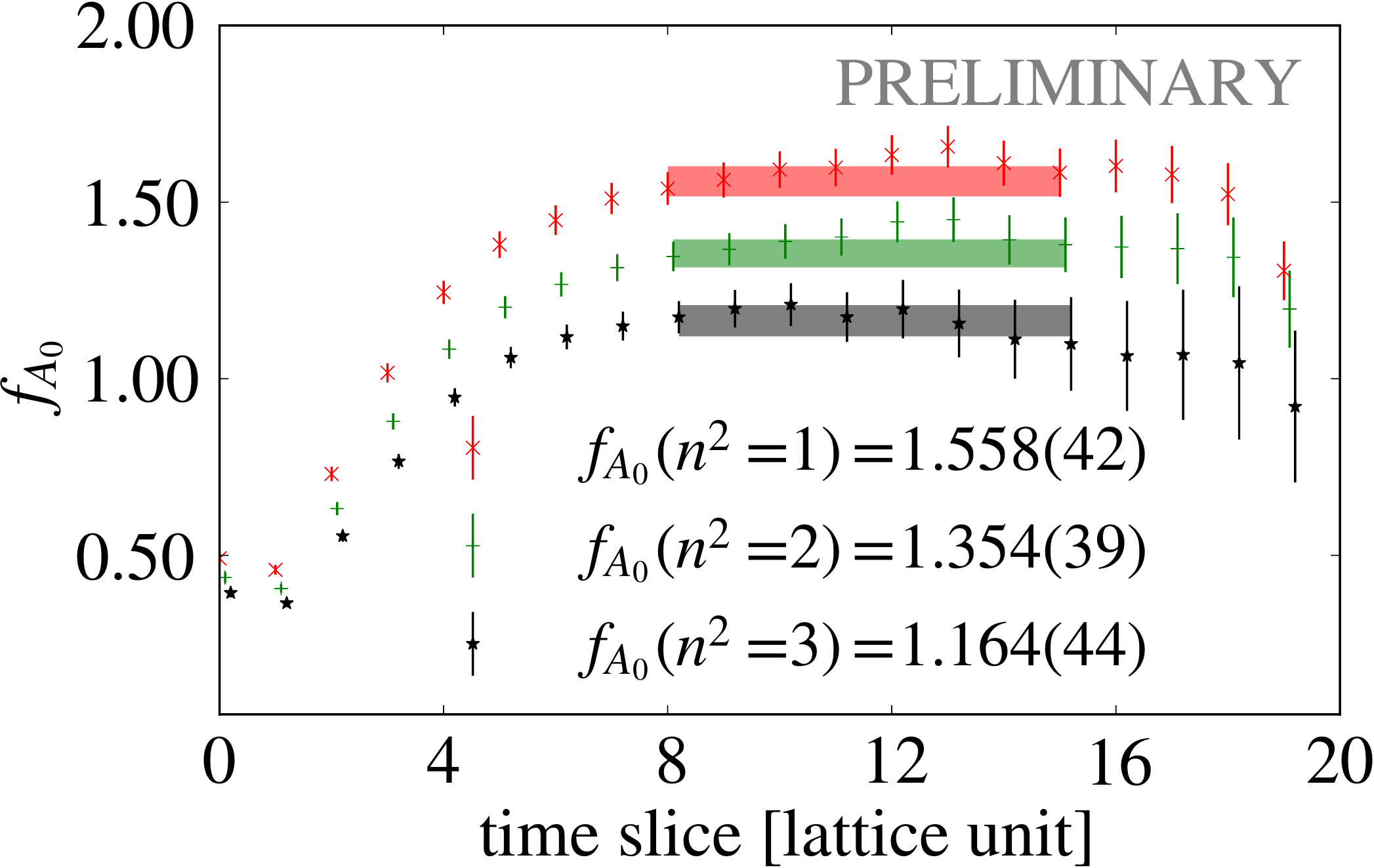}}\vspace{2mm}
\parbox{0.495\textwidth}{\includegraphics[width=\linewidth]{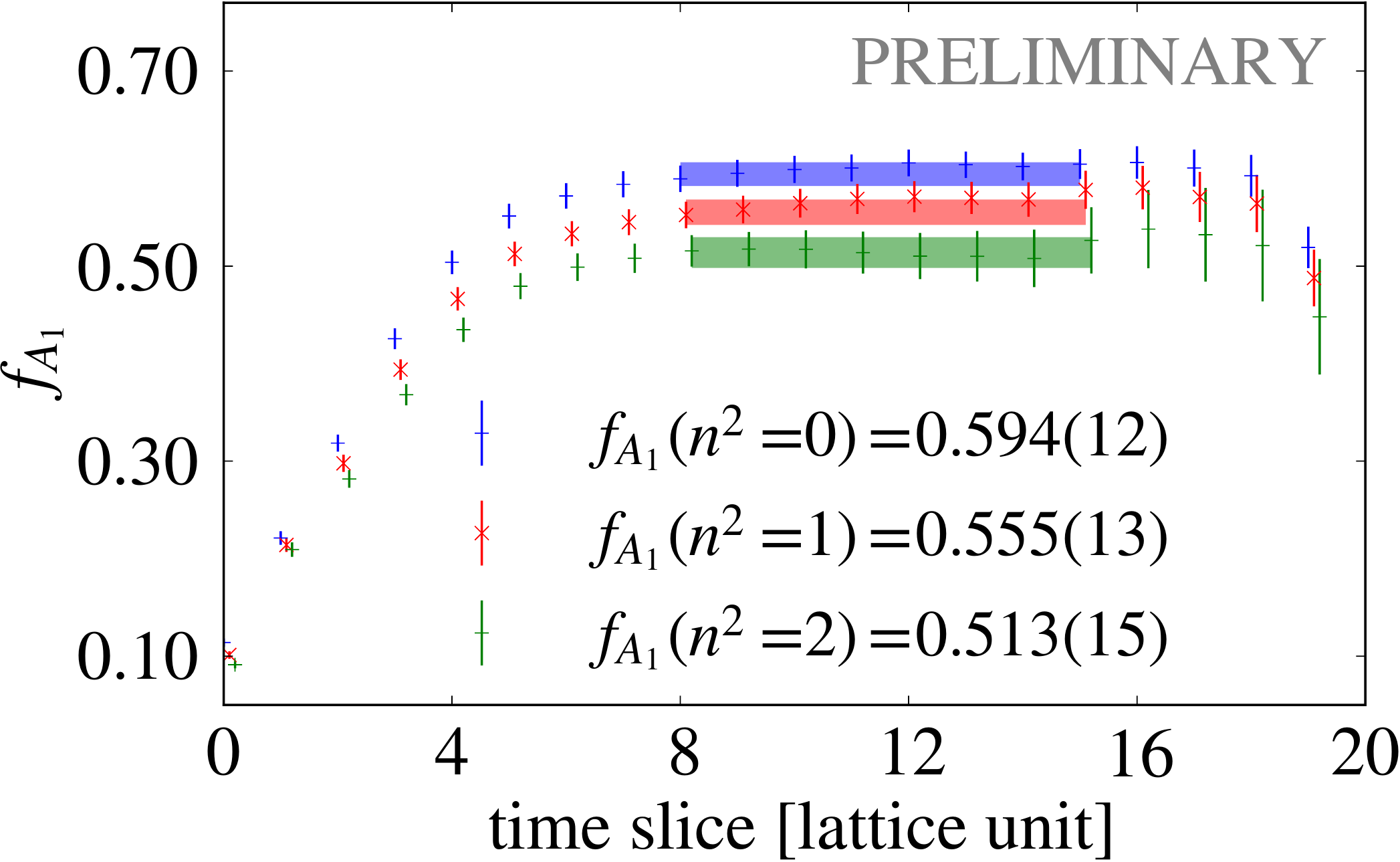}}
\parbox{0.495\textwidth}{\includegraphics[width=\linewidth]{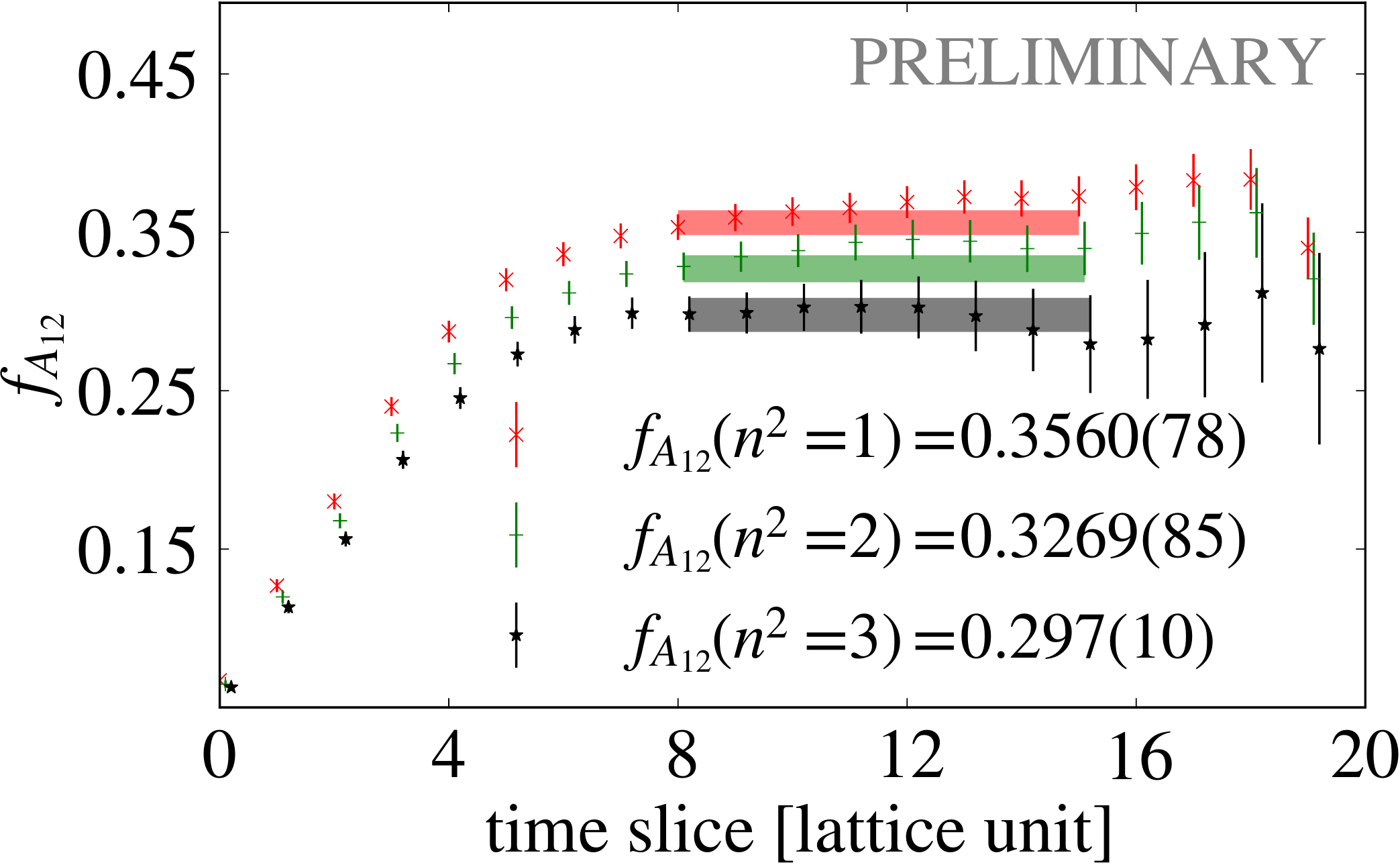}}\vspace{2mm}
\parbox{0.495\textwidth}{\includegraphics[width=\linewidth]{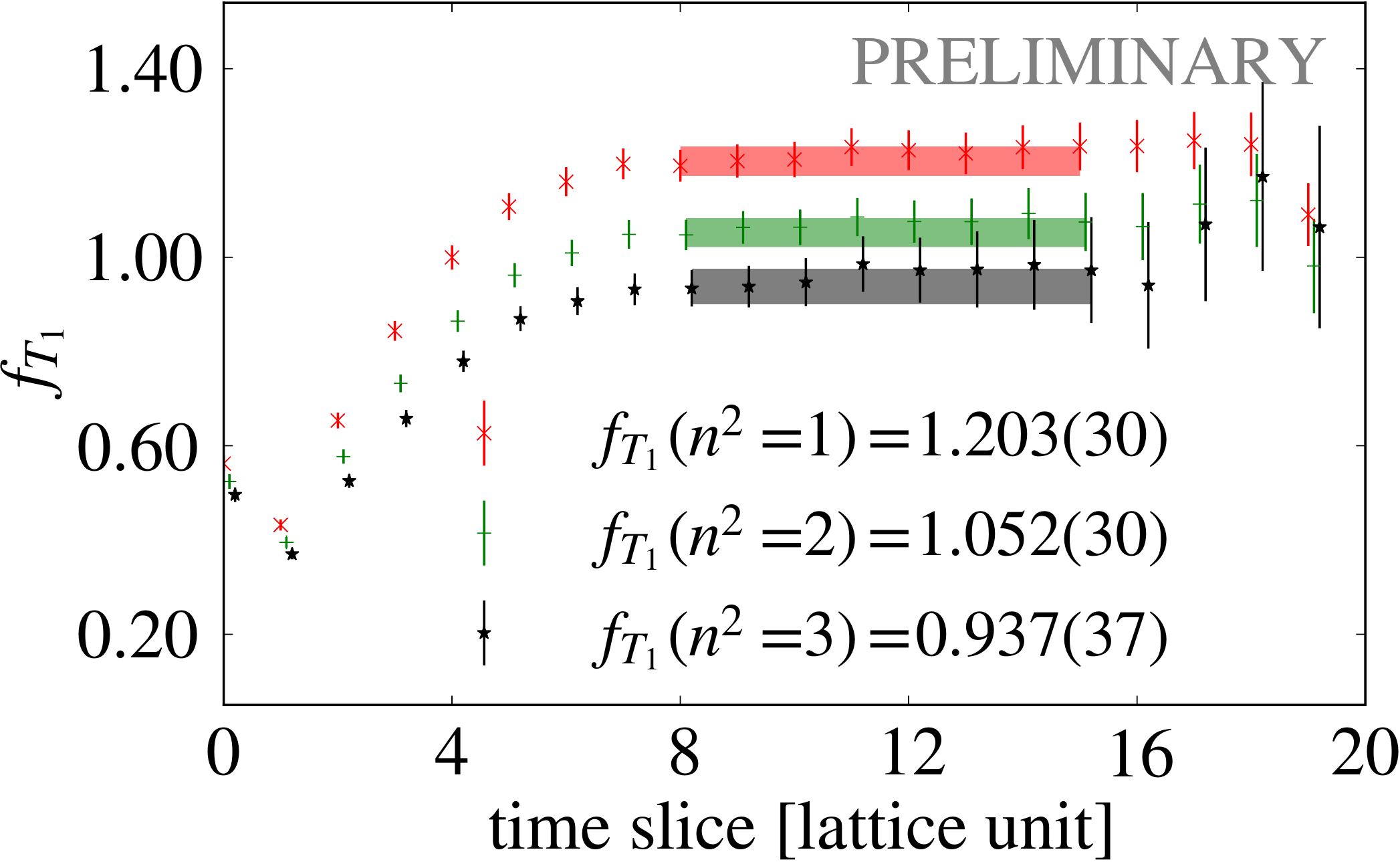}}
\parbox{0.495\textwidth}{\includegraphics[width=\linewidth]{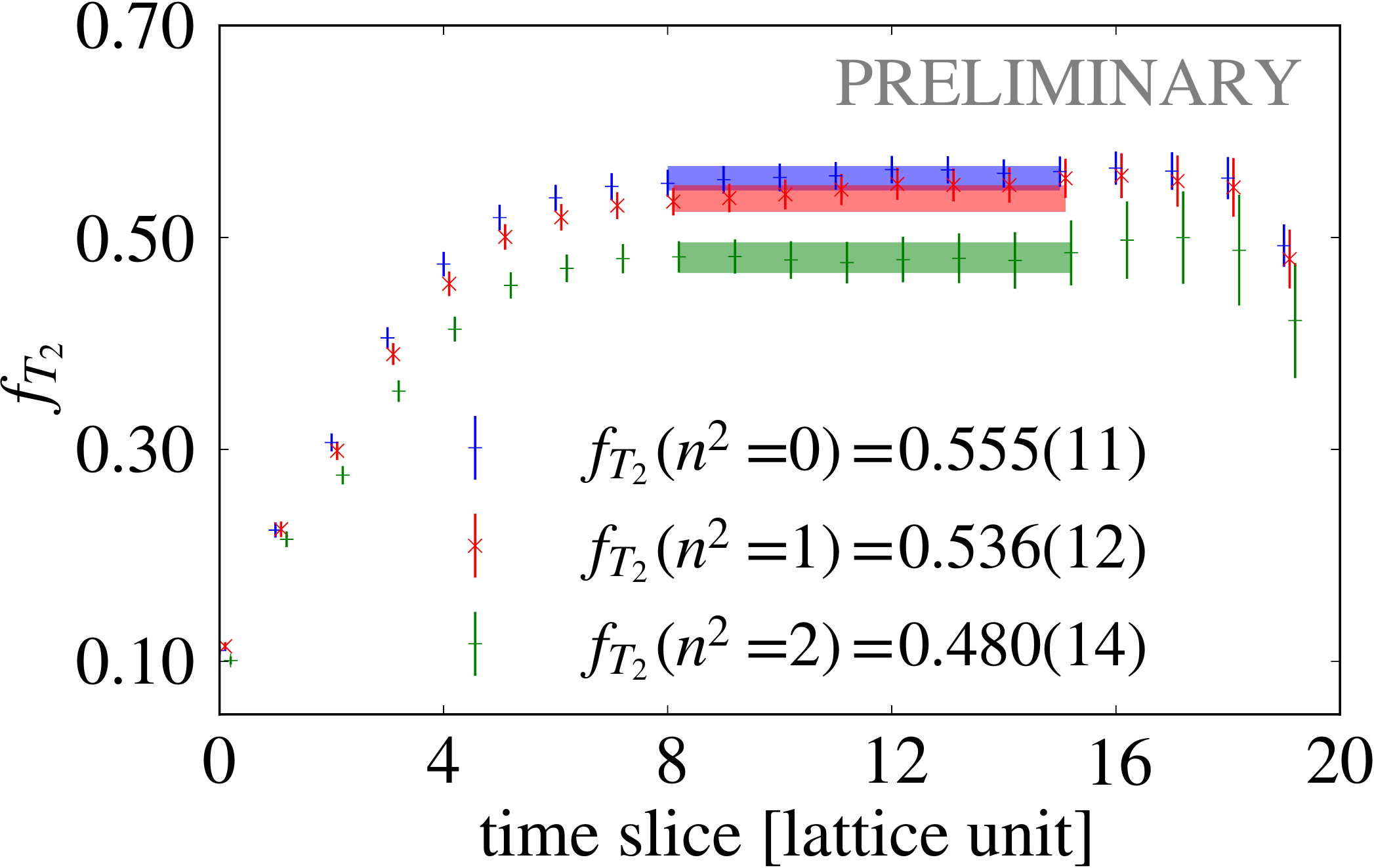}}\vspace{2mm}
\parbox{0.495\textwidth}{\includegraphics[width=\linewidth]{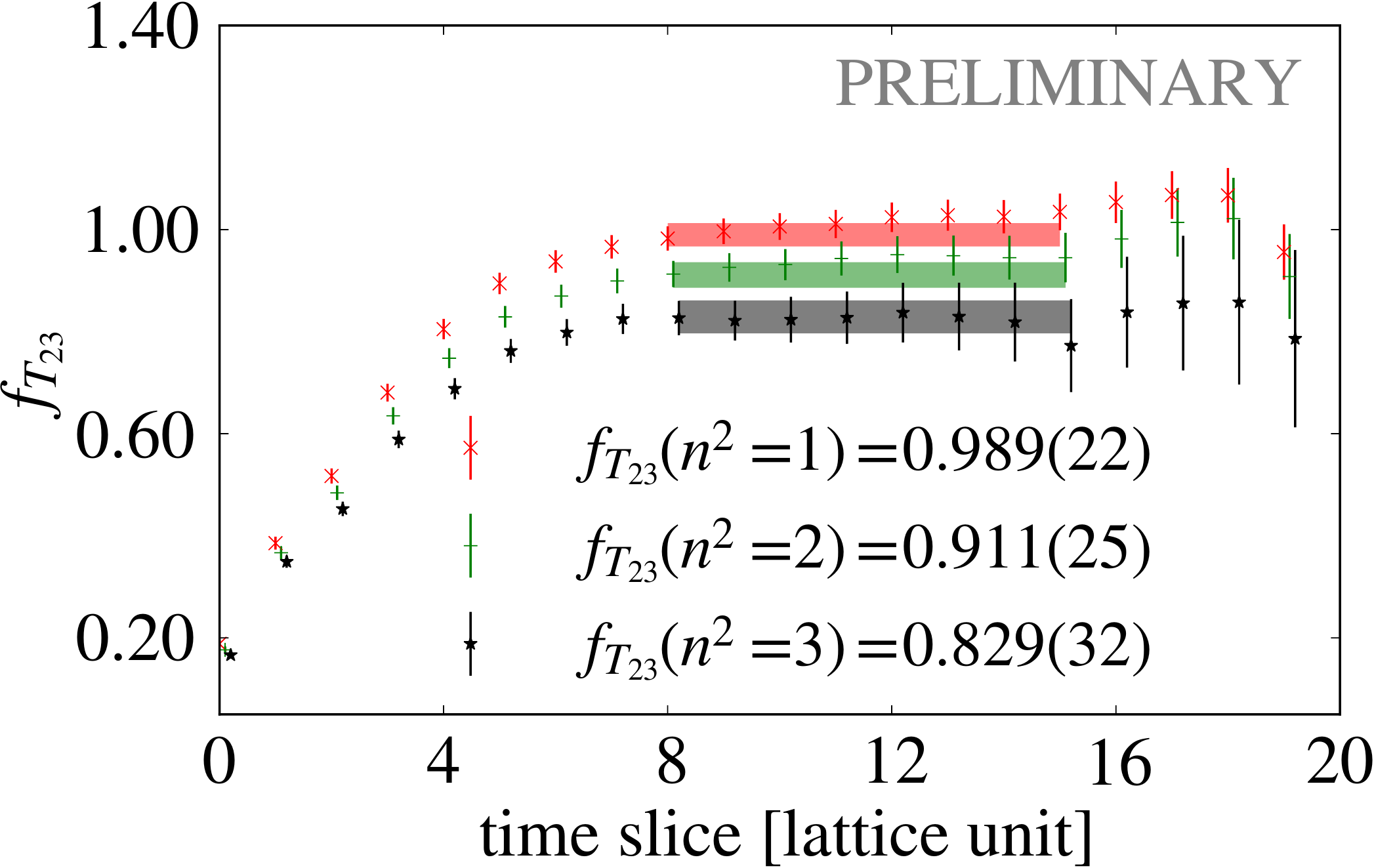}}\hfill ~
\caption{Effective mass style plots showing the determination of the seven form factors $f_V$, $f_{A_0}$,$f_{A_1}$, $f_{A_{12}}$, $f_{T_1}$, $f_{T_2}$ and $f_{T_{23}}$ for $B_s \to \phi \ell^+\ell^-$ for the coarse ensemble ($a^{-1}=1.785$ GeV) with $am_l=0.005$ and $t_\text{sink}-t_\text{source}=20$. We extract the values from correlated, constant in time fits to appropriate linear combinations of ratios of 3-point over 2-point functions. Fitting ranges are indicated by the length of the shaded error band.}
\label{fig:Bs_to_phi_ff_v_t}
\end{figure}
\begin{figure}[p]
\centering
\parbox{0.495\textwidth}{\includegraphics[width=\linewidth]{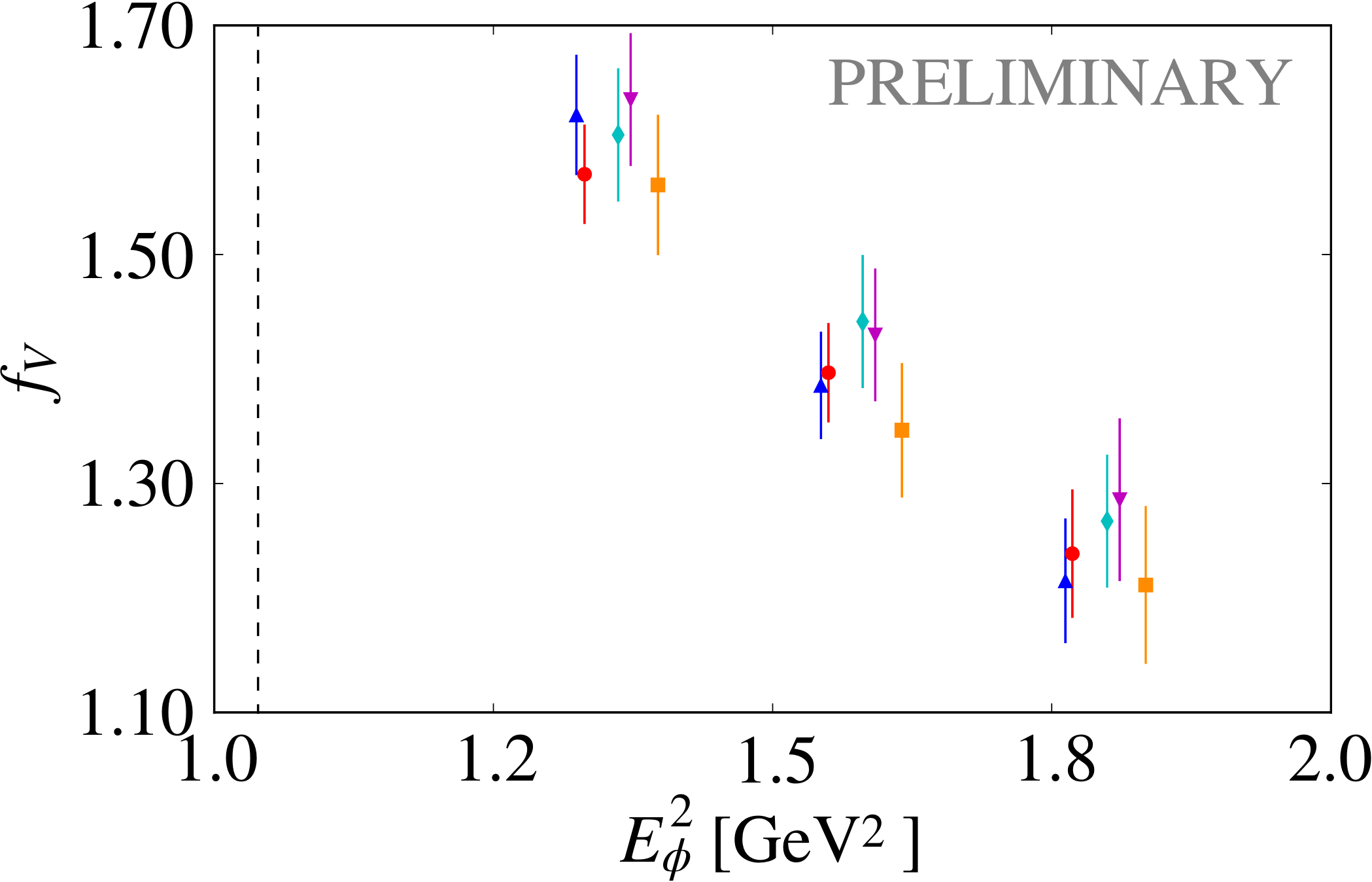}}
\parbox{0.495\textwidth}{\includegraphics[width=\linewidth]{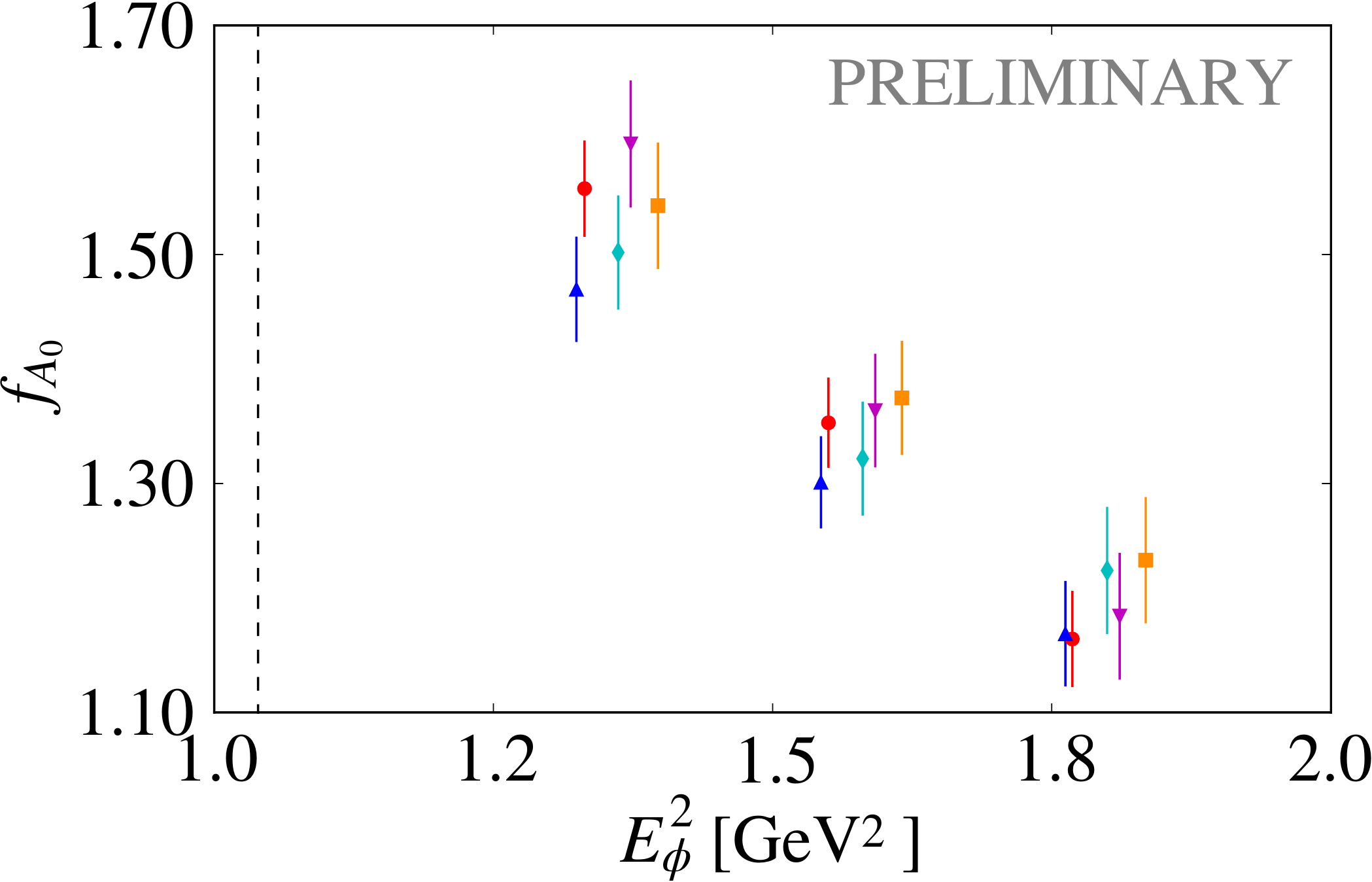}}\vspace{2mm}
\parbox{0.495\textwidth}{\includegraphics[width=\linewidth]{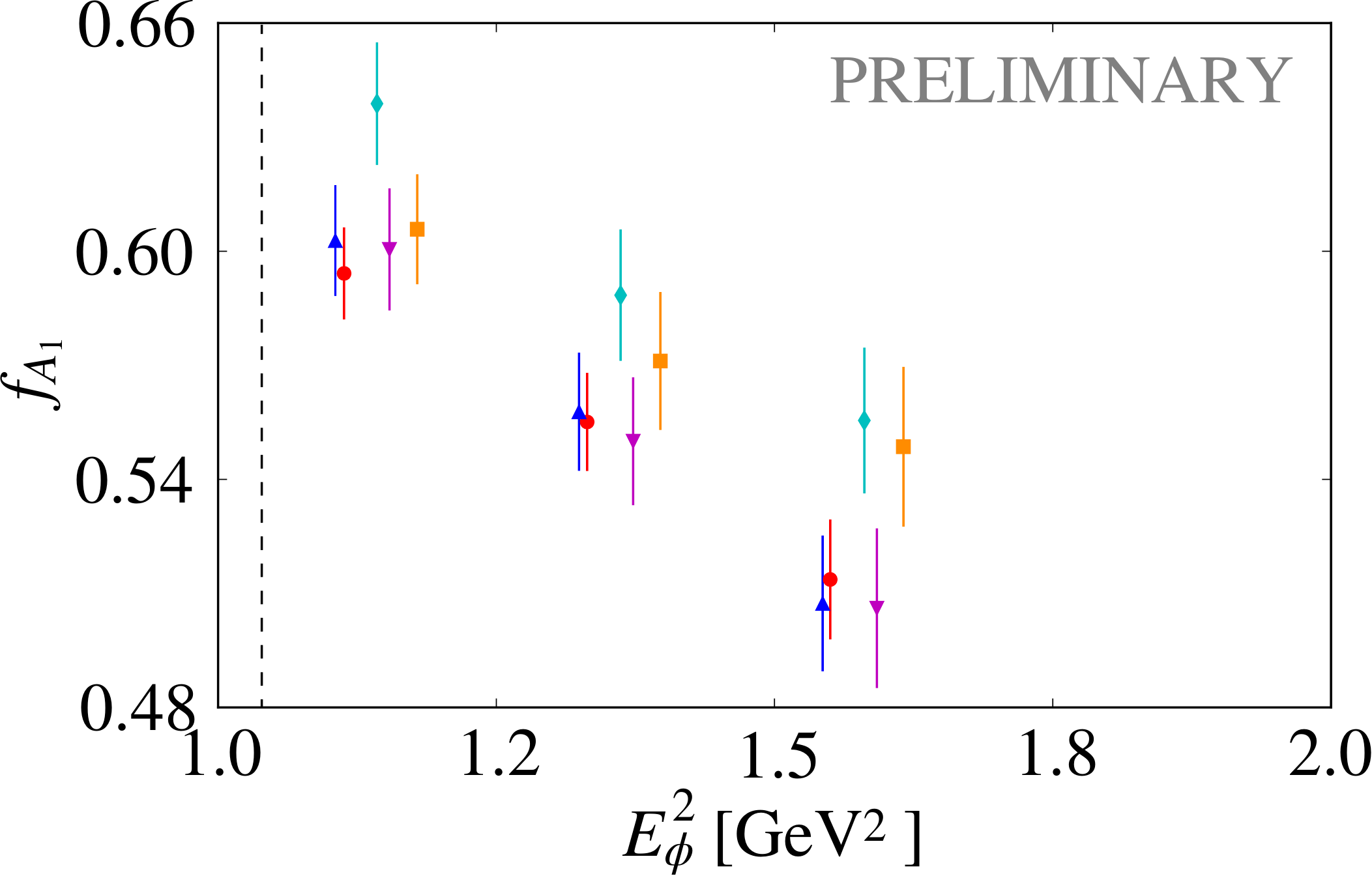}}
\parbox{0.495\textwidth}{\includegraphics[width=\linewidth]{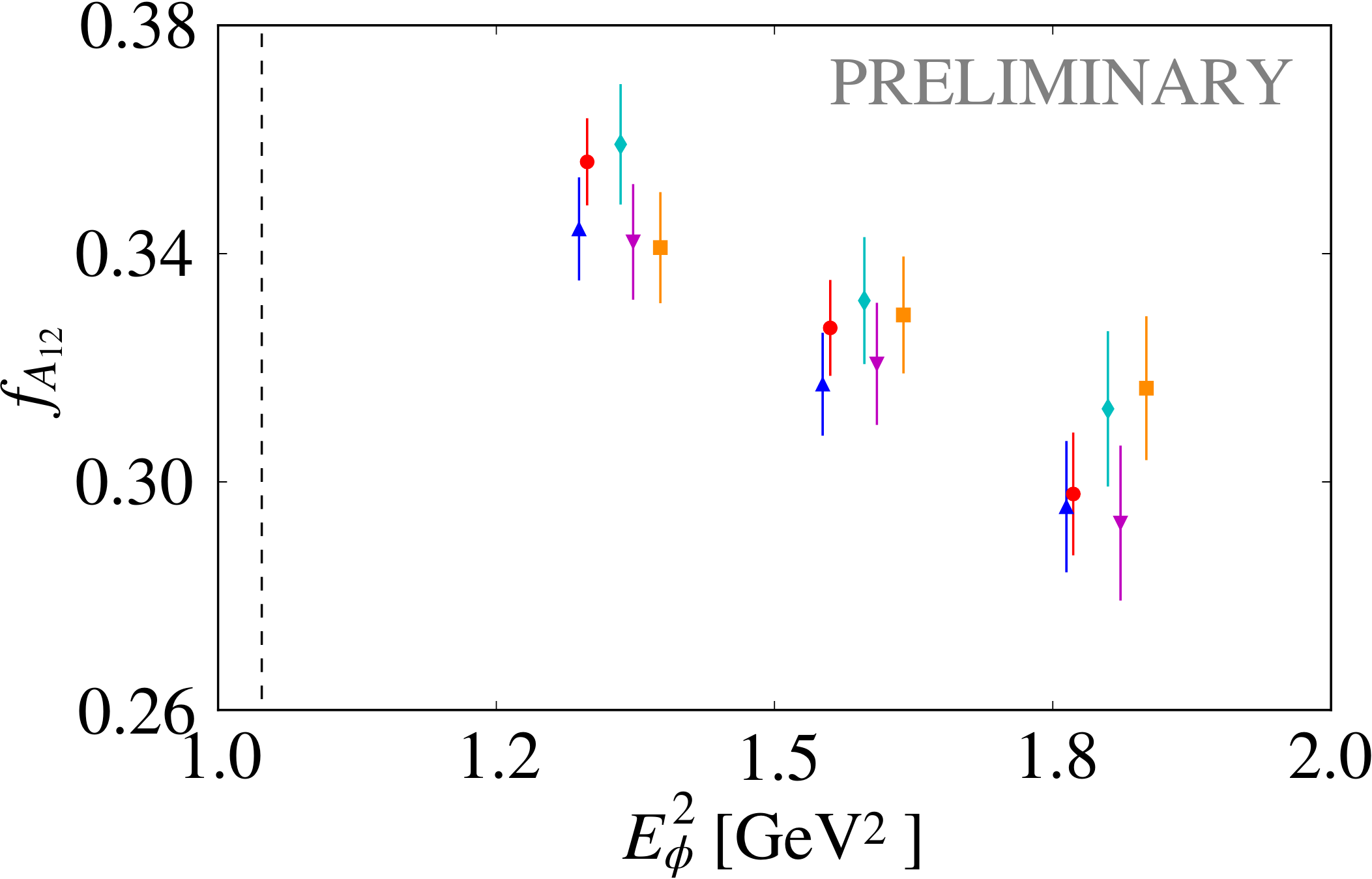}}\vspace{2mm}
\parbox{0.495\textwidth}{\includegraphics[width=\linewidth]{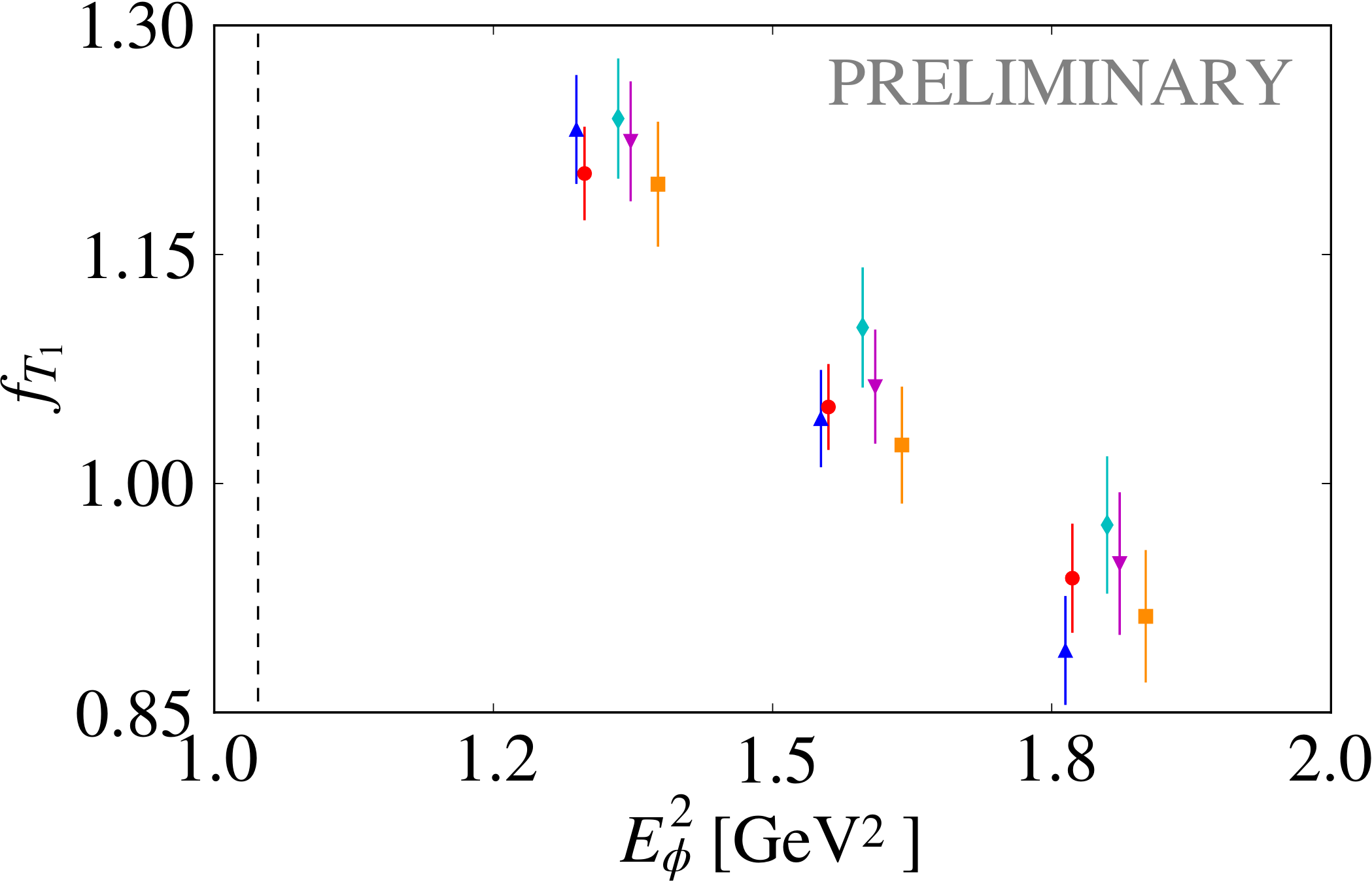}}
\parbox{0.495\textwidth}{\includegraphics[width=\linewidth]{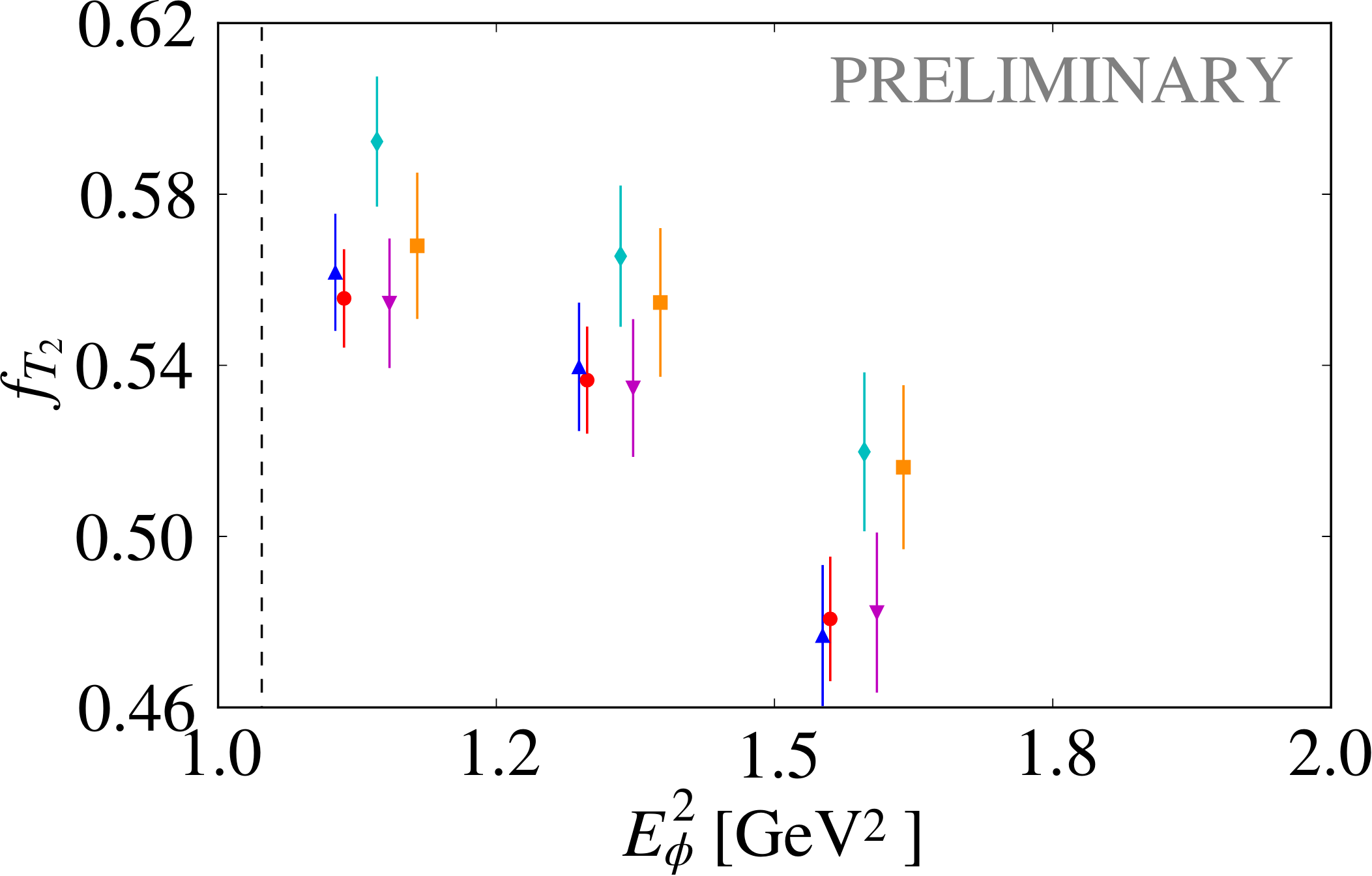}}\vspace{2mm}
\parbox{0.495\textwidth}{\includegraphics[width=\linewidth]{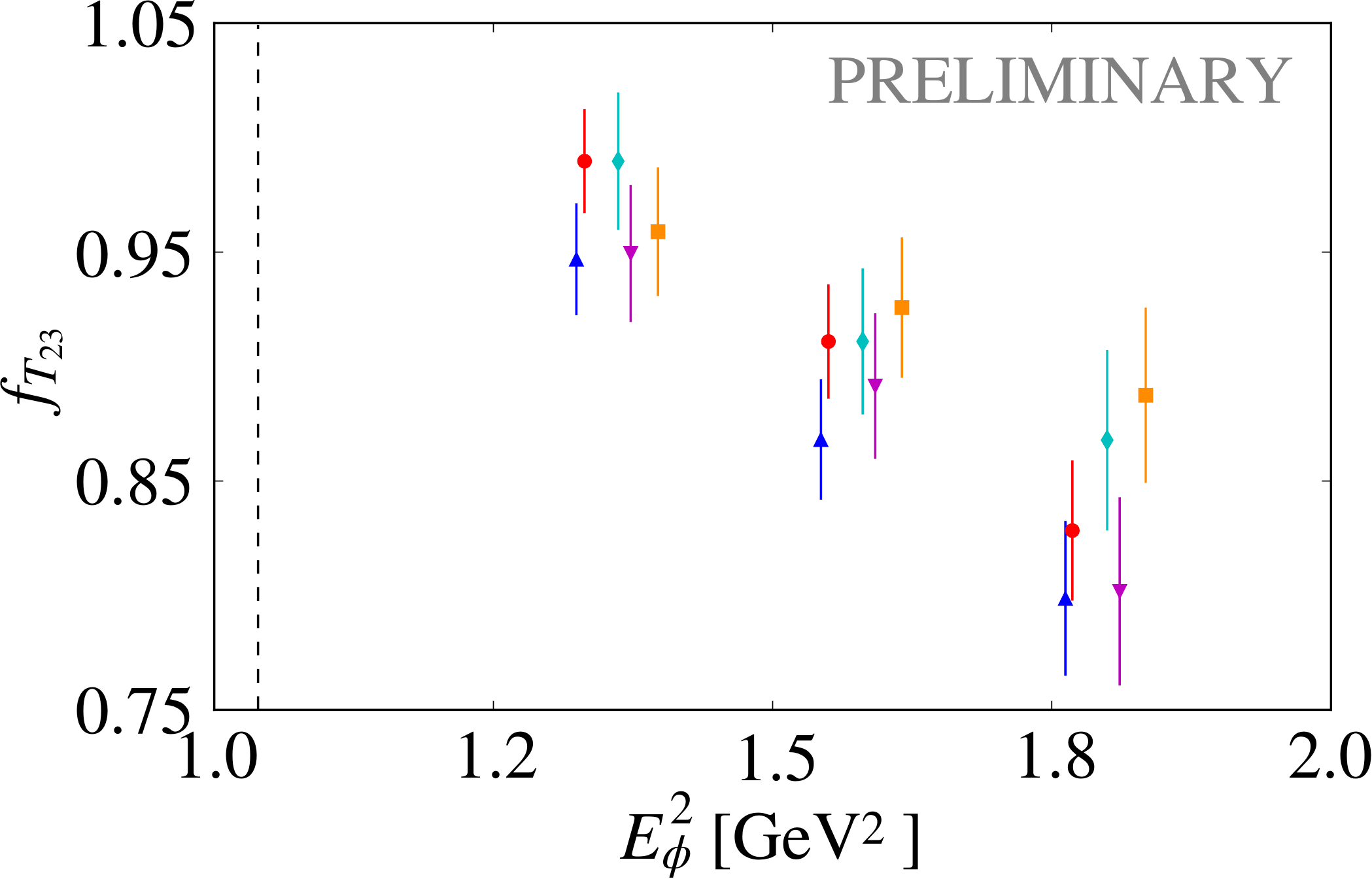}}
\parbox{0.495\textwidth}{\hspace{10mm}\includegraphics[width=0.8\linewidth]{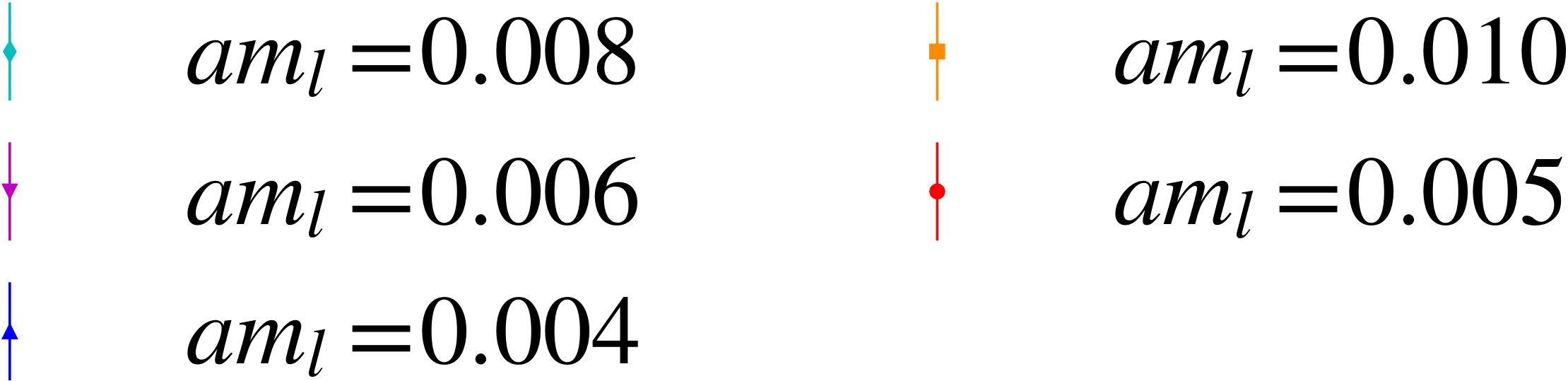}}
\caption{Renormalized results for the seven form factors $f_V, f_{A_0},f_{A_1},
f_{A_{12}}, f_{T_1}, f_{T_2}$ and $f_{T_{23}}$ for $B_s \to \phi \ell^+\ell^-$
versus the squared energy of the hadronic final state.  Shown data are obtained on five different ensembles: sea-quark masses $am_l$ = 0.005, 0.010 correspond to the coarse lattice spacing $a^{-1}=1.785$ GeV and $am_l$ = 0.004, 0.006, 0.008 to the medium fine lattice spacing of $a^{-1}=2.383$ GeV. The dashed lines indicate the physical $\phi$ mass.}
\label{fig:Bs_to_phi}
\end{figure}

%
%
%
The conventional parametrization of $b\to s$ matrix elements is given by a set of seven form factors $f_V$, $f_{A_0}$, $f_{A_1}$, $f_{A_2}$, $f_{T_1}$, $f_{T_2}$ and $f_{T_3}$ \cite{Altmannshofer:2008dz}:
\begin{linenomath*}
  \begin{align}
    \langle \phi( k, \lambda) | \bar{s} \gamma^\mu b | {B_{s}}(p) \rangle &= {f_\phi(q^2)} \frac{2i\epsilon^{\mu\nu\rho\sigma} \varepsilon_{\nu}^*  k_\rho p_\sigma}{M_{B_{s}} + M_\phi}\;, \label{eq:1}\\
    \langle \phi( k, \lambda) | \bar{s} \gamma^\mu \gamma_5 b | {B_{s}}(p) \rangle &= {f_{A_0}(q^2)}\frac{2M_\phi \varepsilon^*\cdot q}{q^2}   q^\mu \notag\\
    &\quad + {f_{A_1} (q^2)}(M_{B_{s}} + M_\phi)\left[ \varepsilon^{*\mu}  - \frac{\varepsilon^*\cdot q}{q^2}   q^\mu \right] \notag\\
    &\quad -{f_{A_2}(q^2)} \frac{\varepsilon^*\cdot q}{M_{B_{s}}+M_\phi} \left[ k^\mu + p^\mu - \frac{M_{B_{s}}^2 -M_\phi^2}{q^2}q^\mu\right] \;,\\
    q_\nu	 \langle \phi( k, \lambda) | \bar{s} \sigma^{\nu\mu} b | {B_{s}}(p) \rangle &= 2{f_{T_1}(q^2)} \epsilon^{\mu\rho\tau\sigma} \varepsilon_{\rho}^* k_\tau p_\sigma \;,\\
    q_\nu \langle \phi( k, \lambda) | \bar{s} \sigma^{\nu\mu} \gamma^5 b | {B_{s}}(p) \rangle &= i {f_{T_2}(q^2)} \left[ \varepsilon^{*\mu} (M_{B_{s}}^2- M_\phi^2) -  (\varepsilon^* \cdot q)(p + k)^\mu \right] \notag\\
    &\quad+ i{f_{T_3} (q^2)}  (\varepsilon^* \cdot q)\left[  q^\mu - \frac{q^2}{M_{B_{s}}^2 - M_\phi^2} (p + k)^\mu \right] \;.
    \label{eq:2}
  \end{align}
\end{linenomath*}
In Equations \eqref{eq:1}-\eqref{eq:2}, the 4-momenta of the $B_s$ 
and $\phi$ mesons are given by $p$ and $k$, respectively, $M_{B_s}$ and $M_\phi$ denote 
the corresponding meson masses, and $q = p-k$ is the momentum transfer. The calculation is carried out in the $B_s$-meson rest frame i.e.~$q=(M_{B_s} - E_{\phi}(|\vec{k}|), - \vec{k})$ . The helicity and polarization vector of the 
$\phi$ meson are denoted by $\lambda$ and $\varepsilon$, respectively. 
The matrix elements in \eqref{eq:1}-\eqref{eq:2} are obtained from the ratio 
\begin{linenomath*}
\begin{align}
 R_{B_{s} \to \phi}^{\alpha\Gamma} (t,t_\text{sink},{k}) 
 &=  \frac{C_{B_{s}\to \phi}^{\alpha\Gamma}(t,t_\text{sink},{k}) }
 {\sqrt{\frac{1}{3}\sum_i C^{ii}_{\phi}(t,{k}) C_{B_{s}} (t_\text{sink}-t)}}
\sqrt{\frac{4E_\phi M_{B_{s}}\sum_\lambda \varepsilon^j(k,\lambda)\varepsilon^{j*}(k,\lambda)}
{e^{-E_\phi t} e^{-M_{B_{s}}(t_\text{sink}-t)}}} \label{eq:ratio}\\
&\xrightarrow{t\to\infty,\;t_\text{sink}-t \to \infty} \sum_\lambda  \varepsilon^\alpha(k, \lambda) \langle \phi(k, \lambda) |  \bar{q}\Gamma b| B_{s}(p) \rangle \;,\label{eq:amplitude}
\end{align}
\end{linenomath*}
with the 3-point functions   
\begin{linenomath*}
\begin{equation}
C^{\alpha\Gamma}_{B_{s}\to \phi}(t, t_\text{sink}, \vec{k}) = 
\sum_{\vec{x},\vec{y}}e^{i\vec{k}\cdot\vec{y}}\langle
\phi(0,\vec{0}; \lambda)\bar{q}(t,\vec{y})\Gamma b(t,\vec{y})B_{s}(t_\text{sink},\vec{x})\rangle .
\label{eq.3pt}
\end{equation}
\end{linenomath*}
We sketch the 3-point functions in Fig.~\ref{fig:3pt} and calculate them by contracting a sequential 
$b$-quark with  a strange quark propagator via a vector or tensor current
$\bar{q}\Gamma b$ with $\Gamma = \{\gamma^\mu, \gamma^5\gamma^\mu, \sigma^{\mu\nu}, \gamma^5\sigma^{\mu\nu}\}$. 
The result is  projected onto states of discrete momenta $\vec k$. The
amplitudes obtained from  Eq.~\eqref{eq:amplitude} allow the straightforward  extraction 
of the form factors $f_{A_0}, f_{A_1}, f_{T_1}$ and $f_{T_2}$. 
We extract the form
factors $f_{A_{12}}$ and $f_{T_{23}}$ following the Cambridge group procedure
Ref.~\cite{Horgan:2013hoa}
\begin{linenomath*} 
\begin{align}   
f_{A_{12}}(q^2) &= \frac{\sqrt{q^2}|\vec{k}|}{8M_{B_s}E_\phi
k_m}\epsilon^*_{0,\mu}R_{B_{s} \to \phi}^{m\gamma^\mu\gamma^5},
\\
f_{T_{23}}(q^2) &= i \frac{|\vec{k}|(M_{B_s} + M_\phi)}{4E_\phi k_m \sqrt{q^2}M_{B_s}}
\epsilon^*_{0,\mu}q_\nu R_{B_{s} \to \phi}^{m\sigma^{\mu\nu}\gamma^5},
\end{align}
\end{linenomath*} 
where 
\begin{equation}
\epsilon^*_{0,\mu} = \frac{1}{\sqrt{q^2}}\left(|\vec{k}|, (E_\phi - M_{B_s})\frac{\vec{k}}{|\vec{k}|}\right). 
\end{equation}    
For our basis of form factors $f_V$, $f_{A_0}$, $f_{A_1}$, $f_{A_{12}}$, $f_{T_1}$, $f_{T_2}$ and $f_{T_{23}}$, we perform 
correlated, constant in time fits up to discretized momenta of $\vec k =2\pi(1,
1, 1)/L$. Within our fitting ranges contamination from excited states is not
visible and we use the same
fitting ranges for all momenta and ensembles at the same lattice spacing. Fitting 
ranges for the $32^3$ ensembles are obtained by scaling our choices on $24^3$ using the ratio of the lattice spacings. 
The resulting form factors are then renormalized following the mostly
non-perturbative method introduced in \cite{Hashimoto:1999yp, ElKhadra:2001rv}
\begin{align}
  \langle \phi(k,\lambda)|{\cal J} | B_s(p)\rangle = \rho^{bs}_{J}
\sqrt{Z_J^{ss} Z_J^{bb}} \langle
\phi(k)|J | B_s(p)\rangle,
\end{align}
where $\cal{J}$ and $J$ are the continuum and lattice current operators,
respectively. We determine the flavor-conserving renormalizaton factors  nonperturbatively in the chiral limit ($Z^{ss}_J = Z^{ll}_J$) and compute $\rho^{bs}_J$ at one loop in mean-field improved lattice perturbation theory for $J = \{\gamma^\mu, \gamma^\mu\gamma^5\}$. For tensorial currents we set the
$\rho^{bs}_J$ factor to its tree-level value because to date
no perturbative one-loop calculation has been pursued. Our preliminary results
for the seven renormalized form factors are presented in
Figs.~\ref{fig:Bs_to_phi_ff_v_t} and \ref{fig:Bs_to_phi}. In Figure \ref{fig:Bs_to_phi_ff_v_t} we show plots of the form factors for our coarse ensemble with $am_l=0.005$ using fixed $t_\text{sink}-t_\text{source}=20$ and show the dependence on the time slices $t$ in between. The error bands show the values for specific momenta extracted from a correlated, constant in time fit.  Figure  \ref{fig:Bs_to_phi} shows the form factors versus the squared energy of the hadronic final state ($\phi$) indicating the physical $\phi$ mass by the dashed line on the left.

\section{Semi-leptonic decays with $b\to c$ transitions}\label{Sec.Charming}
In order to determine semi-leptonic form
factors for bottom quarks transitioning to charm quarks, we need a prescription
for simulating charm quarks on the lattice. Since the mass of the charm quark
($m_c^{\overline{\text{MS}}}(\mu=m_c)=1.27$ GeV \cite{Olive:2016xmw}) is less than our smallest cutoff
($a^{-1}=1.730$ GeV), we may either use an effective action like RHQ or a fully
relativistic formulation based on domain-wall fermions to simulate charm
quarks. While the RHQ action is numerically cheaper, simulating charm with
domain-wall fermions has the advantage that we match the action used for light
and strange quarks. This avoids tuning the three parameters of the RHQ action
and allows us to use a renormalization procedure similar to that in our
 $B\to\pi\ell\nu$ calculation. We therefore simulate charm based
on the recent work featuring optimized M\"obius domain-wall fermions
\cite{Boyle:2015kyy,Boyle:2016imm,Boyle:2016lzk,CharmPaper}
i.e.~we use domain-wall fermions with the M\"obius kernel and choose the
following parameters:
\begin{align}
  L_s &= 12 &\text{(extent of the $5^{th}$ dimension)}\nonumber\\
  M_5&=1.6&\text{(domain-wall height)}\nonumber\\
  b=1.5\quad &\text{and}\quad c=0.5 &\text{(M\"obius parameters)}
\end{align}
without link-smearing the gauge field \cite{Cho:2015ffa,Boyle:2016lzk}. With this set-up, discretization errors  have been
shown to remain small for quantities like the charmonium mass $\eta_c$ or $D_{(s)}$ meson masses and decay constants if bare input quark masses below $am_q \lesssim 0.4$ are chosen \cite{Boyle:2016imm}. Thus on our coarse ensembles
($a^{-1} \approx 1.78$ GeV), we cannot directly simulate charm quarks but expect
a linear extrapolation to be benign \cite{Boyle:2016lzk,CharmPaper}. We simulate 2-3 charm-like quark masses and will subsequently extra-/interpolate to the
physical charm quark mass. The bare charm quark masses used in
our simulations as well as the $\eta_c$ masses relevant for a future extra-/interpolation are listed in Tab.~\ref{tab.charm}.

\begin{table}[t]
  \centering
  \begin{tabular}{cccccccccc} \toprule
    $L^3 \times T$ & $a^{-1}$[GeV] &$am_l$ & $am_h$ & $am_c^\text{sim}$& $\eta_c^\text{sim}$[GeV] \\\midrule
    $24^3 \times 64$ & 1.785(5) & 0.005 & 0.040  & 0.30, 0.35, 0.40 & 2.2246(62),
2.4492(68), 2.6604(74) \\
    $24^3 \times 64$ & 1.785(5) & 0.010 & 0.040 &0.30, 0.35, 0.40&  2.2257(62),
2.4501(68), 2.6612(74) \\ \midrule%
    $32^3 \times 64$ & 2.383(9) & 0.004 & 0.030 &0.28, 0.34&  2.6985(97),
3.059(11)\\
    $32^3 \times 64$ & 2.383(9) & 0.006 & 0.030 &0.28, 0.34&  2.6990(97),
3.059(11)\\
    $32^3 \times 64$ & 2.383(9) & 0.008 & 0.030 &0.28, 0.34&  determination in progress \\
    \bottomrule
  \end{tabular}
\caption{Simulated charm-like bare input quark masses $am_c^\text{sim}$ and the corresponding values of the $\eta_c^\text{sim}$ meson masses in GeV (connected $c\bar c$ contributions only) for our $24^3$ and $32^3$ ensembles. The physical $\eta_c$ mass is $\eta_c^\text{phys}=2.9834(5)$ GeV \cite{Olive:2016xmw}.  }
\label{tab.charm}
\end{table}

\begin{figure}[tb] \centering 
\parbox{0.495\textwidth}{\includegraphics[width=\linewidth]{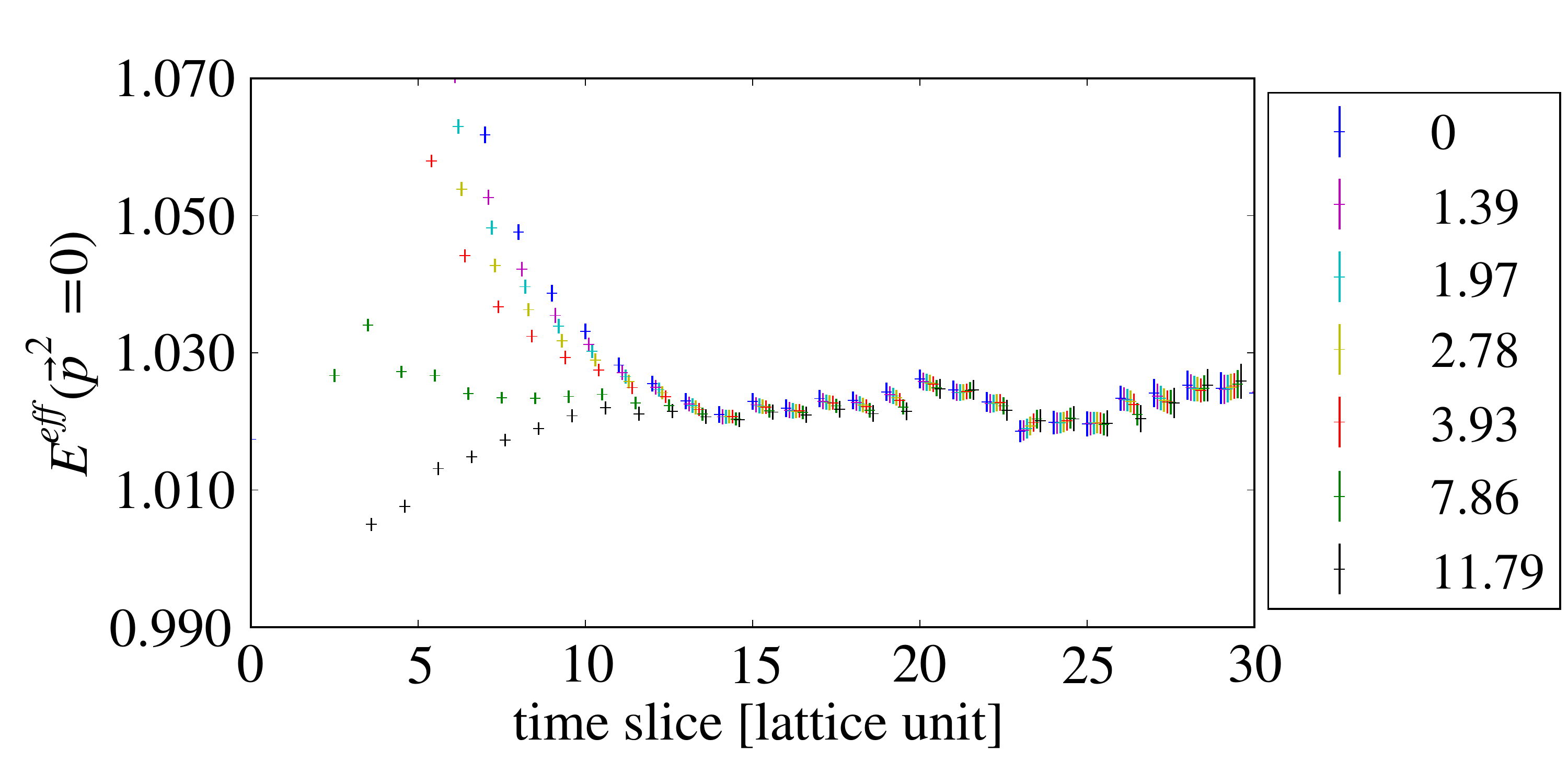}}
\parbox{0.495\textwidth}{\includegraphics[width=\linewidth]{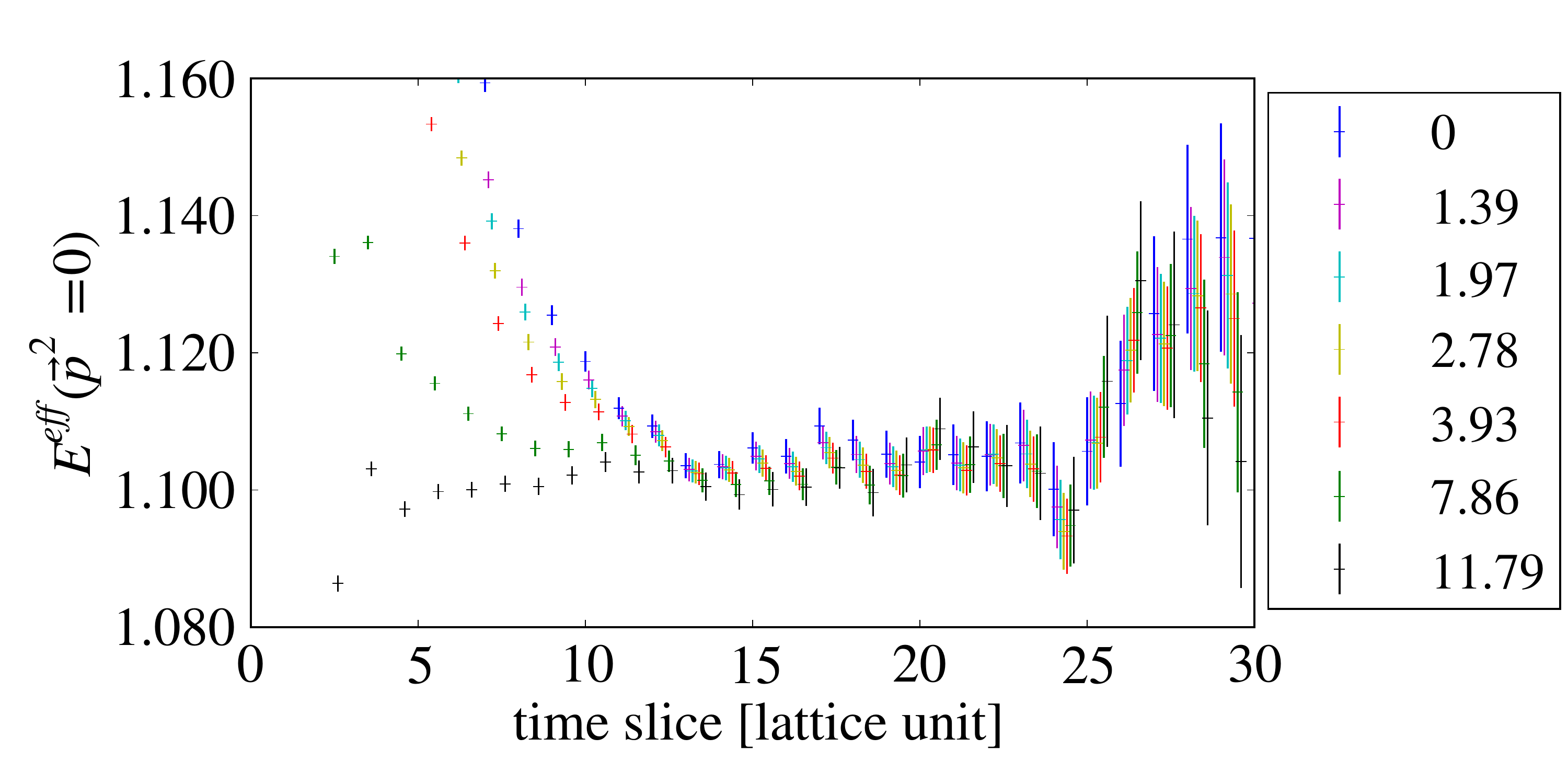}}
\caption{Exploring different widths $\sigma$ of the Gaussian source used to generate charm quarks by comparing effective masses for the $D_s$-like meson on the left and the $D_s^*$-like meson on the right. Strange quarks are generated with a point source and both propagators are contracted with a point sink. The data are obtained on the coarse $24^3$ ensemble with $am_l=0.005$ using $am_s^\text{sim}=0.03224$ and $am_c^\text{sim}=0.400$.}
\label{fig.SourceSmearingCharm}
\end{figure}

Before starting the form factor calculation, we first explored the signal of
charm-light and charm-strange 2-point functions. We repeated a study
investigating Gaussian smeared sources for the charm quarks with different widths $\sigma$ similar to
the one presented in \cite{Aoki:2012xaa}.  In
Figure \ref{fig.SourceSmearingCharm} we show the outcome of this study by
plotting effective masses for $D_s$-like mesons on the left and $D_s^*$-like mesons on
the right obtained on the coarse $24^3$ ensemble with $am_l=0.005$. As can be
seen in the plots, the green data corresponding to a width $\sigma = 7.86$ and $N_\text{smear}=100$ Jacobi iterations result in the earliest onset of the plateau which also extends over many
time slices. Incidentally this is the same outcome as we found in our study for
bottom quarks. Hence we will use the same choice for the Gaussian smearing in
the following. Likewise we verified that the same separation of source and sink
is suitable and results in a good plateau when analyzing the 3-point functions.

In the following we restrict ourselves to $D_{(s)}$ pseudoscalar final states
and introduce the standard form factors $f_+$ and $f_0$ for semileptonic pseudoscalar-to-pseudoscalar decays
\begin{align}
  \langle D_{(s)}(k)|\bar{c}\gamma_\mu b| B_{(s)}(p)\rangle = f_+(q^2)\left(p_\mu + k_\mu - \frac{M^2_{B_{(s)}} - M^2_{D_{(s)}}}{q^2}\right) + f_0(q^2)\frac{M^2_{B_{(s)}} - M^2_{D_{(s)}}}{q^2}q_\mu,
  \label{eq:BtoD}
\end{align}
where $p$ is the 4-momentum of the $B_{(s)}$ meson, $k$ the
4-momentum of the $D_{(s)}$ meson, and $q\equiv (p - k)$, the momentum
transferred to the outgoing charged-lepton-neutrino pair. As before we carry
out our calculation in the $B_{(s)}$-meson rest frame i.e.~$q = \left(
M_{B_{(s)}} - E_{D_{(s)}}(|\vec k|),\;\; -\vec k\right)$.

The form factors introduced in Eq.~(\ref{eq:BtoD}) are defined in the continuum
and related by the bottom-charm renormalization factor $Z_{V_\mu}^{bc}$ to the
matrix elements we determine on the lattice
\begin{align}
  \langle D_{(s)}(k)|{\cal V}_\mu | B_{(s)}(p)\rangle = Z_{V_\mu}^{bc} \langle D_{(s)}(k)|V_\mu | B_{(s)}(p)\rangle.
\end{align}
The continuum vector current operator is denoted by ${\cal V}_\mu = \bar{c}\gamma_\mu b$ and $V_\mu$
is the corresponding lattice current operator. Again we obtain the renormalization
factor by rewriting it according to the mostly non-perturbative method
\cite{Hashimoto:1999yp,ElKhadra:2001rv}
\begin{align}
  Z_{V_\mu}^{bc} = \rho^{bc}_{V_\mu} \sqrt{Z_V^{cc} Z_V^{bb}},
\end{align}
where the flavor conserving factors are
determined non-perturbatively and only the remaining $\rho$-factor is
determined using lattice perturbation theory.

\begin{figure}[t]
  \centering 
\parbox{0.495\textwidth}{\includegraphics[width=\linewidth]{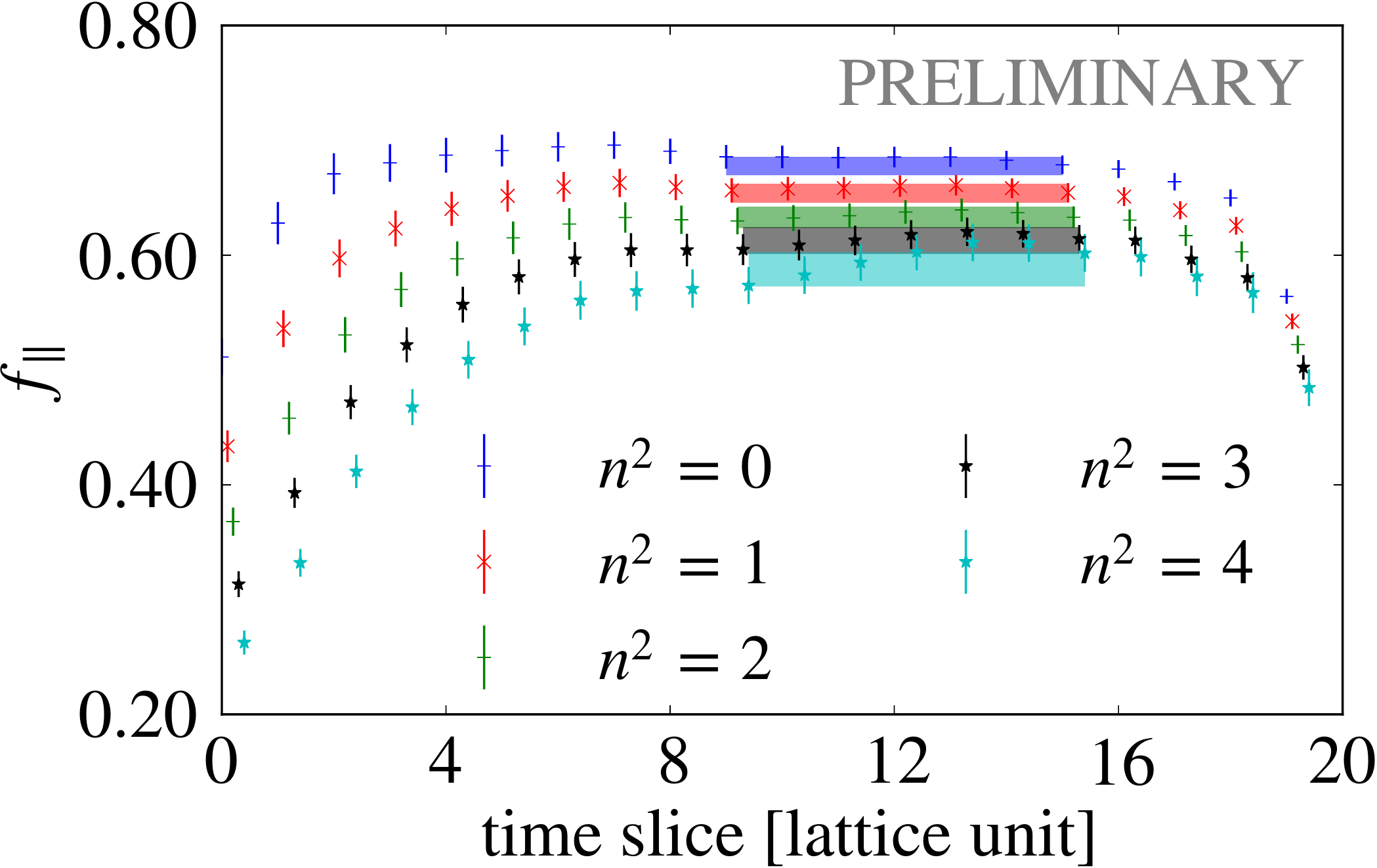}}
\parbox{0.495\textwidth}{\includegraphics[width=\linewidth]{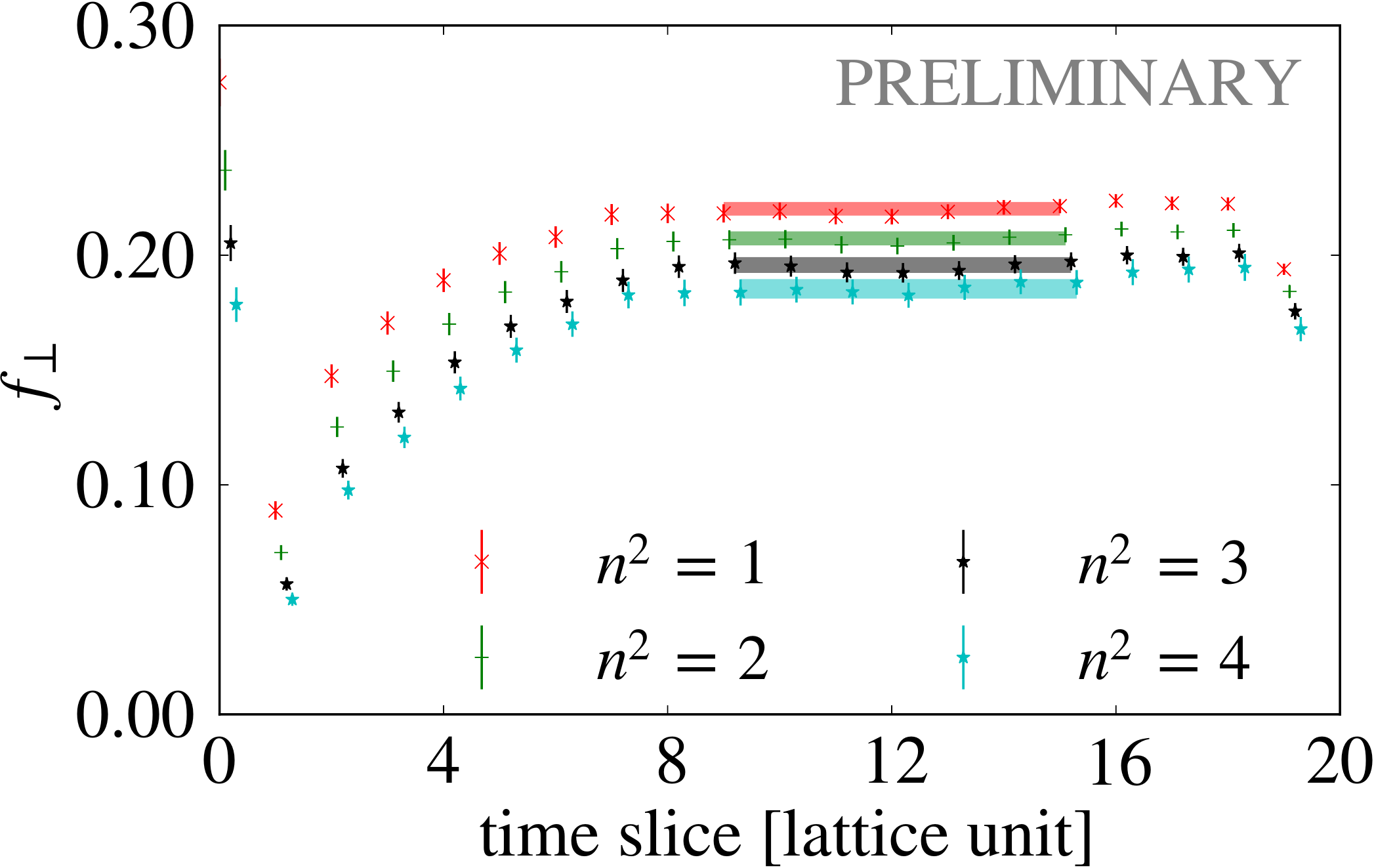}}\vspace{2mm}
\parbox{0.495\textwidth}{\includegraphics[width=\linewidth]{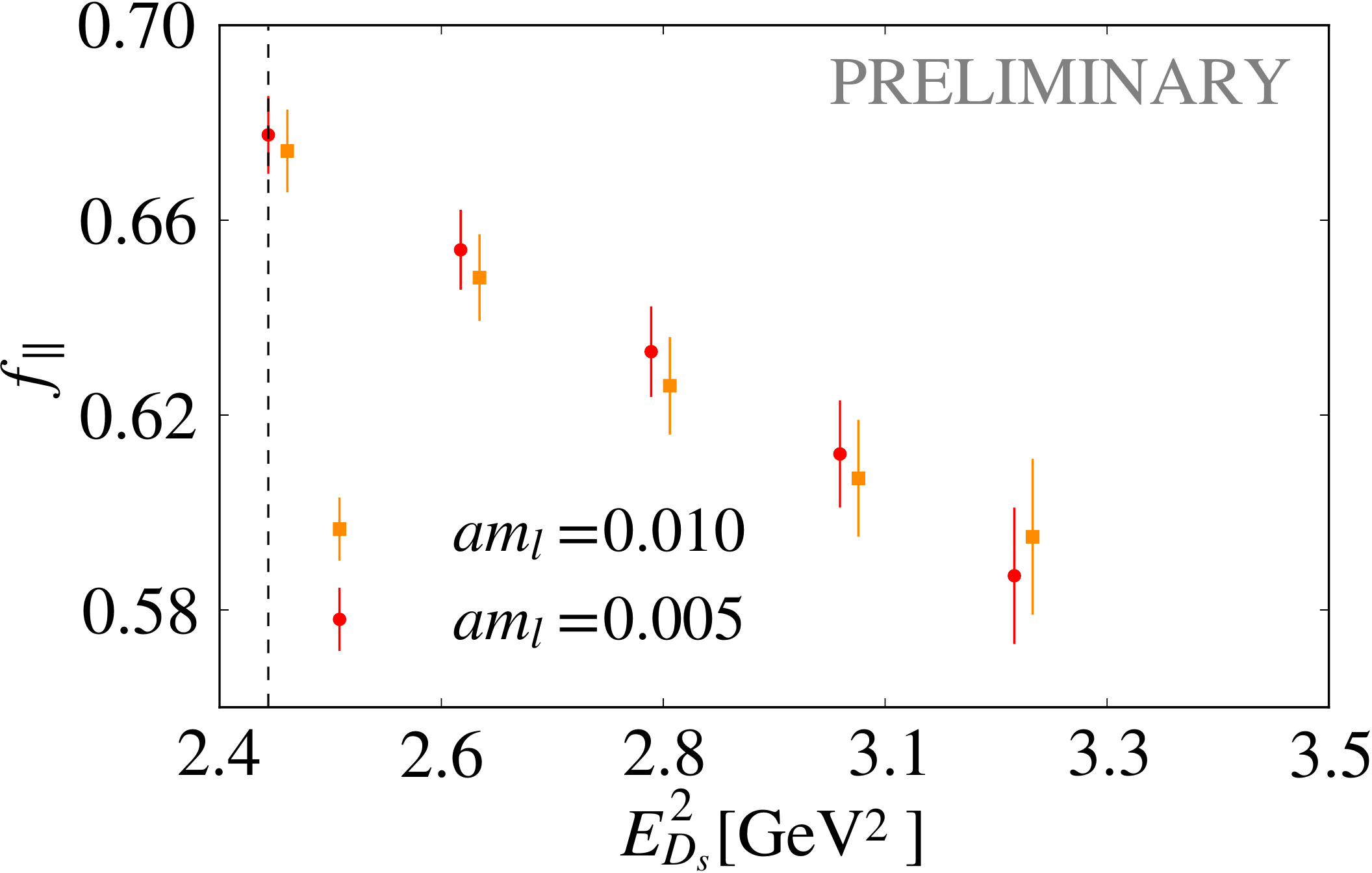}}
\parbox{0.495\textwidth}{\includegraphics[width=\linewidth]{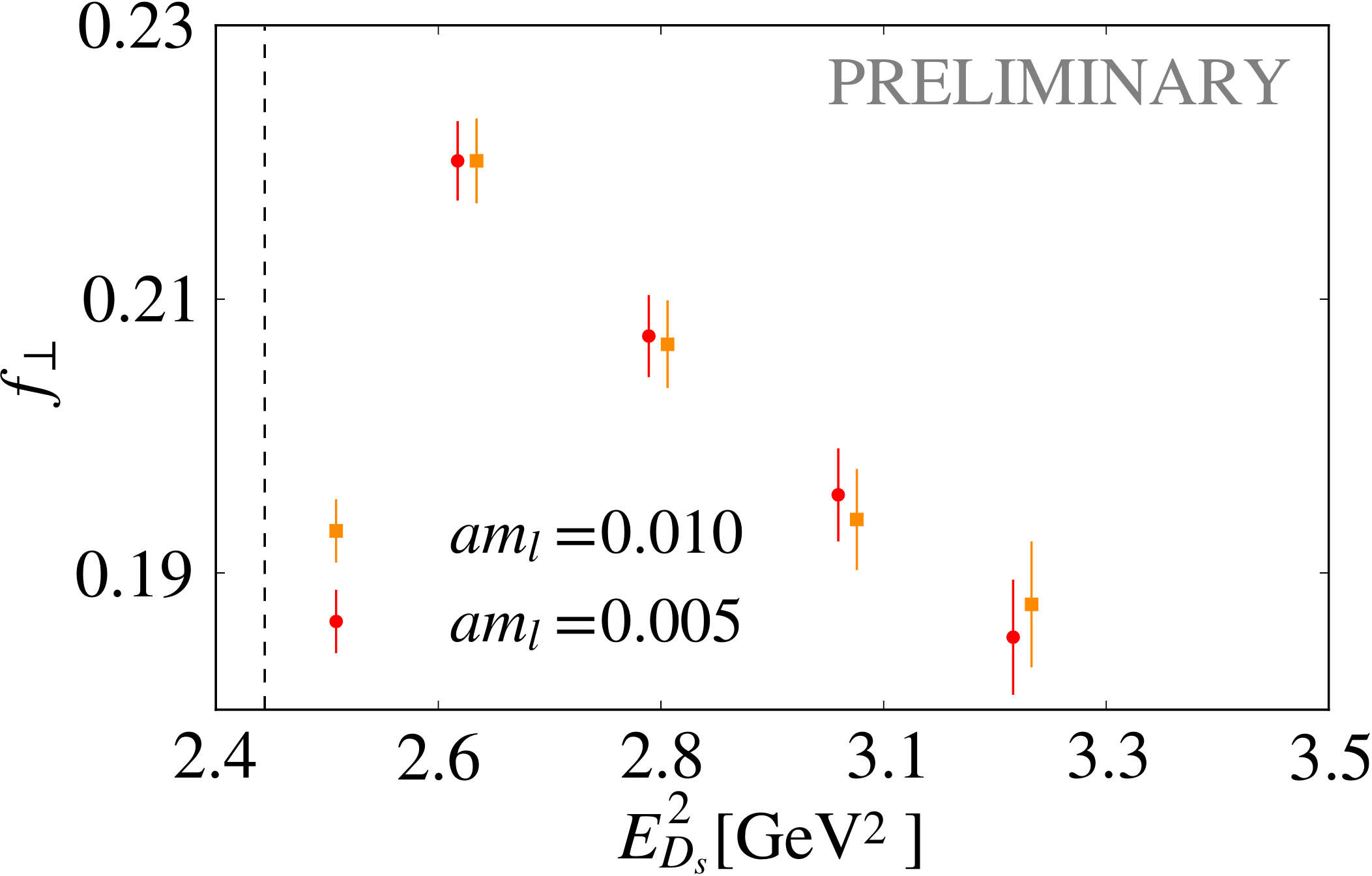}}
\caption{Preliminary results for our determination of semi-leptonic form
factors for $B_s\to D_s\ell\nu$. On the left we show $f_\parallel$, on the
right $f_\perp$.  The upper plots show fits to the linear combination of ratios
of 3-point over 2-point functions determining the form factor for the coarse
ensemble ($a^{-1}=1.785$ GeV) with $am_l=0.005$ using $am_c^\text{sim}=0.300$.
In the lower plots we show our results for both coarse ensembles as a function of the squared energy of the $D_s$-like meson. The mass of the $D_s$-like meson is indicated by the black dashed line.}
\label{fig.BstoDs}
\end{figure}

Following the prescription given in Refs.~\cite{Boyle:2016mwo,Boyle:2016wis}, $Z_V^{cc}$ is obtained in the chiral limit using the domain-wall height $M_5$ chosen to simulate charm quarks. In Figure \ref{fig.BstoDs} we show our preliminary results for $B_s \to D_s \ell\nu$ semi-leptonic decays in terms of the form factors $f_\parallel$ and $f_\perp$ obtained on the $24^3$ ensembles with $a^{-1}=1.785$ GeV. $f_\parallel$ and $f_\perp$ are linearly related to the physical form factors $f_+$ and $f_0$. These data have yet to be renormalized and are based on the tree-level operator only. We show discretized momenta up to $\vec k = 2\pi\vec n/L$  with $\vec n$ up to $(2,0,0)$ and average spatial directions with the same $|n|$. \vspace{-2mm}

\section{Outlook and conclusions}
\label{Sec.Conclusion} 
We have reported on the status of our program to compute semi-leptonic $B$ decays. The full program considers pseudoscalar  $B$ or $B_s$ mesons in the initial state and a single pseudoscalar or vector  meson in the final state. Vector final states are treated as stable within the narrow width approximation. Here we show updates for our calculation of GIM suppressed $B_s\to \phi \ell^+ \ell^-$ decays and presented first results for computing $b\to c$ transitions as they occur e.g.~in $B_s\to D_s\ell \nu$.

Most of our numerical simulations have been completed on the $24^3$ and $32^3$ ensembles, but we are still accumulating data on the more costly $48^3$ ensembles. In parallel we are carrying out the perturbative calculations required for renormalizing and $O(a)$-improving the weak matrix elements calculated on the lattice and start to build-up our data analysis. Once we have $O(a)$-improved and renormalized data we will start to carry out combined chiral- and continuum extrapolations and for $B_{(s)}\to D_{(s)}^{(*)}\ell \nu$ decays will explore extra-/interpolating to the physical charm quark mass.

\paragraph{Acknowledgments} The authors thank our collaborators in the RBC and
UKQCD Collaborations for helpful discussions and suggestions.  Computations for
this work were performed on resources provided by the USQCD Collaboration,
funded by the Office of Science of the U.S.~Department of Energy, as well as on
computers at Columbia University and Brookhaven National Laboratory.
This work used the ARCHER UK National Supercomputing Service (\href{http://www.archer.ac.uk}{http://www.archer.ac.uk}).
Gauge field configurations on which our calculations are based were also generated
using the DiRAC Blue Gene Q system at the University of Edinburgh, part of the
DiRAC Facility; funded by BIS National E-infrastructure grant ST/K000411/1 and
STFC grants ST/H008845/1, ST/K005804/1 and ST/K005790/1.  This project has
received funding from the European Union's Horizon 2020 research and innovation
programme under the Marie Sk{\l}odowska-Curie grant agreement No 659322, the
European Research Council under the European Unions Seventh Framework Programme
(FP7/ 2007-2013) / ERC Grant agreement 279757, STFC grant ST/L000296/1 and
ST/L000458/1 as well as the EPSRC Doctoral Training Centre grant
(EP/G03690X/1). No new experimental data was generated for this research.

{\small \bibliography{B_meson} \bibliographystyle{apsrev4-1} }

\begin{thebibliography}{52}%
\makeatletter
\providecommand \@ifxundefined [1]{%
 \@ifx{#1\undefined}
}%
\providecommand \@ifnum [1]{%
 \ifnum #1\expandafter \@firstoftwo
 \else \expandafter \@secondoftwo
 \fi
}%
\providecommand \@ifx [1]{%
 \ifx #1\expandafter \@firstoftwo
 \else \expandafter \@secondoftwo
 \fi
}%
\providecommand \natexlab [1]{#1}%
\providecommand \enquote  [1]{``#1''}%
\providecommand \bibnamefont  [1]{#1}%
\providecommand \bibfnamefont [1]{#1}%
\providecommand \citenamefont [1]{#1}%
\providecommand \href@noop [0]{\@secondoftwo}%
\providecommand \href [0]{\begingroup \@sanitize@url \@href}%
\providecommand \@href[1]{\@@startlink{#1}\@@href}%
\providecommand \@@href[1]{\endgroup#1\@@endlink}%
\providecommand \@sanitize@url [0]{\catcode `\\12\catcode `\$12\catcode
  `\&12\catcode `\#12\catcode `\^12\catcode `\_12\catcode `\%12\relax}%
\providecommand \@@startlink[1]{}%
\providecommand \@@endlink[0]{}%
\providecommand \url  [0]{\begingroup\@sanitize@url \@url }%
\providecommand \@url [1]{\endgroup\@href {#1}{\urlprefix }}%
\providecommand \urlprefix  [0]{URL }%
\providecommand \Eprint [0]{\href }%
\providecommand \doibase [0]{http://dx.doi.org/}%
\providecommand \selectlanguage [0]{\@gobble}%
\providecommand \bibinfo  [0]{\@secondoftwo}%
\providecommand \bibfield  [0]{\@secondoftwo}%
\providecommand \translation [1]{[#1]}%
\providecommand \BibitemOpen [0]{}%
\providecommand \bibitemStop [0]{}%
\providecommand \bibitemNoStop [0]{.\EOS\space}%
\providecommand \EOS [0]{\spacefactor3000\relax}%
\providecommand \BibitemShut  [1]{\csname bibitem#1\endcsname}%
\let\auto@bib@innerbib\@empty
\bibitem [{\citenamefont {Glashow}\ \emph {et~al.}(1970)\citenamefont
  {Glashow}, \citenamefont {Iliopoulos},\ and\ \citenamefont
  {Maiani}}]{Glashow:1970gm}%
  \BibitemOpen
  \bibfield  {author} {\bibinfo {author} {\bibfnamefont {S.~L.}\ \bibnamefont
  {Glashow}}, \bibinfo {author} {\bibfnamefont {J.}~\bibnamefont {Iliopoulos}},
  \ and\ \bibinfo {author} {\bibfnamefont {L.}~\bibnamefont {Maiani}},\ }\href
  {\doibase 10.1103/PhysRevD.2.1285} {\bibfield  {journal} {\bibinfo  {journal}
  {Phys. Rev.}\ }\textbf {\bibinfo {volume} {D2}},\ \bibinfo {pages} {1285}
  (\bibinfo {year} {1970})}
\bibitem [{\citenamefont {Aaij}\ \emph {et~al.}(2016)\citenamefont {Aaij} \emph
  {et~al.}}]{Aaij:2015oid}%
  \BibitemOpen
  \bibfield  {author} {\bibinfo {author} {\bibfnamefont {R.}~\bibnamefont
  {Aaij}} \emph {et~al.} (\bibinfo {collaboration} {LHCb}),\ }\href {\doibase
  10.1007/JHEP02(2016)104} {\bibfield  {journal} {\bibinfo  {journal} {JHEP}\
  }\textbf {\bibinfo {volume} {02}},\ \bibinfo {pages} {104} (\bibinfo {year}
  {2016})},\ \Eprint {http://arxiv.org/abs/1512.04442} {arXiv:1512.04442
  [hep-ex]}
\bibitem [{\citenamefont {Aaij}\ \emph {et~al.}(2015)\citenamefont {Aaij} \emph
  {et~al.}}]{Aaij:2015esa}%
  \BibitemOpen
  \bibfield  {author} {\bibinfo {author} {\bibfnamefont {R.}~\bibnamefont
  {Aaij}} \emph {et~al.} (\bibinfo {collaboration} {LHCb}),\ }\href {\doibase
  10.1007/JHEP09(2015)179} {\bibfield  {journal} {\bibinfo  {journal} {JHEP}\
  }\textbf {\bibinfo {volume} {09}},\ \bibinfo {pages} {179} (\bibinfo {year}
  {2015})},\ \Eprint {http://arxiv.org/abs/1506.08777} {arXiv:1506.08777
  [hep-ex]}
\bibitem [{\citenamefont {Aaij}\ \emph {et~al.}(2014)\citenamefont {Aaij} \emph
  {et~al.}}]{Aaij:2014ora}%
  \BibitemOpen
  \bibfield  {author} {\bibinfo {author} {\bibfnamefont {R.}~\bibnamefont
  {Aaij}} \emph {et~al.} (\bibinfo {collaboration} {LHCb}),\ }\href {\doibase
  10.1103/PhysRevLett.113.151601} {\bibfield  {journal} {\bibinfo  {journal}
  {Phys. Rev. Lett.}\ }\textbf {\bibinfo {volume} {113}},\ \bibinfo {pages}
  {151601} (\bibinfo {year} {2014})},\ \Eprint {http://arxiv.org/abs/1406.6482}
  {arXiv:1406.6482 [hep-ex]}
\bibitem [{\citenamefont {Fajfer}\ \emph {et~al.}(2012)\citenamefont {Fajfer},
  \citenamefont {Kamenik},\ and\ \citenamefont {Nisandzic}}]{Fajfer:2012vx}%
  \BibitemOpen
  \bibfield  {author} {\bibinfo {author} {\bibfnamefont {S.}~\bibnamefont
  {Fajfer}}, \bibinfo {author} {\bibfnamefont {J.~F.}\ \bibnamefont {Kamenik}},
  \ and\ \bibinfo {author} {\bibfnamefont {I.}~\bibnamefont {Nisandzic}},\
  }\href {\doibase 10.1103/PhysRevD.85.094025} {\bibfield  {journal} {\bibinfo
  {journal} {Phys. Rev.}\ }\textbf {\bibinfo {volume} {D85}},\ \bibinfo {pages}
  {094025} (\bibinfo {year} {2012})},\ \Eprint {http://arxiv.org/abs/1203.2654}
  {arXiv:1203.2654 [hep-ph]}
\bibitem [{\citenamefont {Bailey}\ \emph {et~al.}(2012)\citenamefont {Bailey}
  \emph {et~al.}}]{Bailey:2012jg}%
  \BibitemOpen
  \bibfield  {author} {\bibinfo {author} {\bibfnamefont {J.~A.}\ \bibnamefont
  {Bailey}} \emph {et~al.},\ }\href {\doibase 10.1103/PhysRevLett.109.071802}
  {\bibfield  {journal} {\bibinfo  {journal} {Phys. Rev. Lett.}\ }\textbf
  {\bibinfo {volume} {109}},\ \bibinfo {pages} {071802} (\bibinfo {year}
  {2012})},\ \Eprint {http://arxiv.org/abs/1206.4992} {arXiv:1206.4992
  [hep-ph]}
\bibitem [{\citenamefont {Lees}\ \emph {et~al.}(2012)\citenamefont {Lees} \emph
  {et~al.}}]{Lees:2012xj}%
  \BibitemOpen
  \bibfield  {author} {\bibinfo {author} {\bibfnamefont {J.}~\bibnamefont
  {Lees}} \emph {et~al.} (\bibinfo {collaboration} {BaBar}),\ }\href {\doibase
  10.1103/PhysRevLett.109.101802} {\bibfield  {journal} {\bibinfo  {journal}
  {Phys.Rev.Lett.}\ }\textbf {\bibinfo {volume} {109}},\ \bibinfo {pages}
  {101802} (\bibinfo {year} {2012})},\ \Eprint {http://arxiv.org/abs/1205.5442}
  {arXiv:1205.5442 [hep-ex]}
\bibitem [{\citenamefont {Nandi}\ \emph {et~al.}(2016)\citenamefont {Nandi},
  \citenamefont {Patra},\ and\ \citenamefont {Soni}}]{Nandi:2016wlp}%
  \BibitemOpen
  \bibfield  {author} {\bibinfo {author} {\bibfnamefont {S.}~\bibnamefont
  {Nandi}}, \bibinfo {author} {\bibfnamefont {S.~K.}\ \bibnamefont {Patra}}, \
  and\ \bibinfo {author} {\bibfnamefont {A.}~\bibnamefont {Soni}},\ }\href@noop
  {} {\  (\bibinfo {year} {2016})},\ \Eprint {http://arxiv.org/abs/1605.07191}
  {arXiv:1605.07191 [hep-ph]}
\bibitem [{\citenamefont {Flynn}\ \emph {et~al.}(2015)\citenamefont {Flynn},
  \citenamefont {Izubuchi}, \citenamefont {Kawanai}, \citenamefont {Lehner},
  \citenamefont {Soni}, \citenamefont {Van~de Water},\ and\ \citenamefont
  {Witzel}}]{Flynn:2015mha}%
  \BibitemOpen
  \bibfield  {author} {\bibinfo {author} {\bibfnamefont {J.~M.}\ \bibnamefont
  {Flynn}}, \bibinfo {author} {\bibfnamefont {T.}~\bibnamefont {Izubuchi}},
  \bibinfo {author} {\bibfnamefont {T.}~\bibnamefont {Kawanai}}, \bibinfo
  {author} {\bibfnamefont {C.}~\bibnamefont {Lehner}}, \bibinfo {author}
  {\bibfnamefont {A.}~\bibnamefont {Soni}}, \bibinfo {author} {\bibfnamefont
  {R.~S.}\ \bibnamefont {Van~de Water}}, \ and\ \bibinfo {author}
  {\bibfnamefont {O.}~\bibnamefont {Witzel}},\ }\href {\doibase
  10.1103/PhysRevD.91.074510} {\bibfield  {journal} {\bibinfo  {journal} {Phys.
  Rev.}\ }\textbf {\bibinfo {volume} {D91}},\ \bibinfo {pages} {074510}
  (\bibinfo {year} {2015})},\ \Eprint {http://arxiv.org/abs/1501.05373}
  {arXiv:1501.05373 [hep-lat]}
\bibitem [{\citenamefont {Flynn}\ \emph {et~al.}(2016)\citenamefont {Flynn},
  \citenamefont {Jüttner}, \citenamefont {Kawanai}, \citenamefont {Lizarazo},\
  and\ \citenamefont {Witzel}}]{Flynn:2015ynk}%
  \BibitemOpen
  \bibfield  {author} {\bibinfo {author} {\bibfnamefont {J.}~\bibnamefont
  {Flynn}}, \bibinfo {author} {\bibfnamefont {A.}~\bibnamefont {Jüttner}},
  \bibinfo {author} {\bibfnamefont {T.}~\bibnamefont {Kawanai}}, \bibinfo
  {author} {\bibfnamefont {E.}~\bibnamefont {Lizarazo}}, \ and\ \bibinfo
  {author} {\bibfnamefont {O.}~\bibnamefont {Witzel}},\ }\href@noop {}
  {\bibfield  {journal} {\bibinfo  {journal} {PoS}\ }\textbf {\bibinfo {volume}
  {LATTICE2015}},\ \bibinfo {pages} {345} (\bibinfo {year} {2016})},\ \Eprint
  {http://arxiv.org/abs/1511.06622} {arXiv:1511.06622 [hep-lat]}
\bibitem [{\citenamefont {Bigi}\ and\ \citenamefont
  {Gambino}(2016)}]{Bigi:2016mdz}%
  \BibitemOpen
  \bibfield  {author} {\bibinfo {author} {\bibfnamefont {D.}~\bibnamefont
  {Bigi}}\ and\ \bibinfo {author} {\bibfnamefont {P.}~\bibnamefont {Gambino}},\
  }\href {\doibase 10.1103/PhysRevD.94.094008} {\bibfield  {journal} {\bibinfo
  {journal} {Phys. Rev.}\ }\textbf {\bibinfo {volume} {D94}},\ \bibinfo {pages}
  {094008} (\bibinfo {year} {2016})},\ \Eprint
  {http://arxiv.org/abs/1606.08030} {arXiv:1606.08030 [hep-ph]}
\bibitem [{\citenamefont {Allton}\ \emph {et~al.}(2008)\citenamefont {Allton}
  \emph {et~al.}}]{Allton:2008pn}%
  \BibitemOpen
  \bibfield  {author} {\bibinfo {author} {\bibfnamefont {C.}~\bibnamefont
  {Allton}} \emph {et~al.} (\bibinfo {collaboration} {RBC-UKQCD}),\ }\href
  {\doibase 10.1103/PhysRevD.78.114509} {\bibfield  {journal} {\bibinfo
  {journal} {Phys. Rev.}\ }\textbf {\bibinfo {volume} {D78}},\ \bibinfo {pages}
  {114509} (\bibinfo {year} {2008})},\ \Eprint {http://arxiv.org/abs/0804.0473}
  {arXiv:0804.0473 [hep-lat]}
\bibitem [{\citenamefont {Aoki}\ \emph {et~al.}(2011)\citenamefont {Aoki} \emph
  {et~al.}}]{Aoki:2010dy}%
  \BibitemOpen
  \bibfield  {author} {\bibinfo {author} {\bibfnamefont {Y.}~\bibnamefont
  {Aoki}} \emph {et~al.} (\bibinfo {collaboration} {RBC-UKQCD}),\ }\href
  {\doibase 10.1103/PhysRevD.83.074508} {\bibfield  {journal} {\bibinfo
  {journal} {Phys.Rev.}\ }\textbf {\bibinfo {volume} {D83}},\ \bibinfo {pages}
  {074508} (\bibinfo {year} {2011})},\ \Eprint {http://arxiv.org/abs/1011.0892}
  {arXiv:1011.0892 [hep-lat]}
\bibitem [{\citenamefont {Blum}\ \emph {et~al.}(2016)\citenamefont {Blum} \emph
  {et~al.}}]{Blum:2014tka}%
  \BibitemOpen
  \bibfield  {author} {\bibinfo {author} {\bibfnamefont {T.}~\bibnamefont
  {Blum}} \emph {et~al.} (\bibinfo {collaboration} {RBC, UKQCD}),\ }\href
  {\doibase 10.1103/PhysRevD.93.074505} {\bibfield  {journal} {\bibinfo
  {journal} {Phys. Rev.}\ }\textbf {\bibinfo {volume} {D93}},\ \bibinfo {pages}
  {074505} (\bibinfo {year} {2016})},\ \Eprint {http://arxiv.org/abs/1411.7017}
  {arXiv:1411.7017 [hep-lat]}
\bibitem [{\citenamefont {Boyle}\ \emph
  {et~al.}(2016{\natexlab{a}})\citenamefont {Boyle} \emph
  {et~al.}}]{CharmPaper}%
  \BibitemOpen
  \bibfield  {author} {\bibinfo {author} {\bibfnamefont {P.~A.}\ \bibnamefont
  {Boyle}} \emph {et~al.},\ }\href@noop {} {\enquote {\bibinfo {title} {in
  preparation},}\ } (\bibinfo {year} {2016}{\natexlab{a}})
\bibitem [{\citenamefont {Shamir}(1993)}]{Shamir:1993zy}%
  \BibitemOpen
  \bibfield  {author} {\bibinfo {author} {\bibfnamefont {Y.}~\bibnamefont
  {Shamir}},\ }\href {\doibase 10.1016/0550-3213(93)90162-I} {\bibfield
  {journal} {\bibinfo  {journal} {Nucl. Phys.}\ }\textbf {\bibinfo {volume}
  {B406}},\ \bibinfo {pages} {90} (\bibinfo {year} {1993})},\ \Eprint
  {http://arxiv.org/abs/hep-lat/9303005} {arXiv:hep-lat/9303005}
\bibitem [{\citenamefont {Furman}\ and\ \citenamefont
  {Shamir}(1995)}]{Furman:1994ky}%
  \BibitemOpen
  \bibfield  {author} {\bibinfo {author} {\bibfnamefont {V.}~\bibnamefont
  {Furman}}\ and\ \bibinfo {author} {\bibfnamefont {Y.}~\bibnamefont
  {Shamir}},\ }\href {\doibase 10.1016/0550-3213(95)00031-M} {\bibfield
  {journal} {\bibinfo  {journal} {Nucl. Phys.}\ }\textbf {\bibinfo {volume}
  {B439}},\ \bibinfo {pages} {54} (\bibinfo {year} {1995})},\ \Eprint
  {http://arxiv.org/abs/hep-lat/9405004} {arXiv:hep-lat/9405004}
\bibitem [{\citenamefont {Brower}\ \emph {et~al.}(2012)\citenamefont {Brower},
  \citenamefont {Neff},\ and\ \citenamefont {Orginos}}]{Brower:2012vk}%
  \BibitemOpen
  \bibfield  {author} {\bibinfo {author} {\bibfnamefont {R.~C.}\ \bibnamefont
  {Brower}}, \bibinfo {author} {\bibfnamefont {H.}~\bibnamefont {Neff}}, \ and\
  \bibinfo {author} {\bibfnamefont {K.}~\bibnamefont {Orginos}},\ }\href@noop
  {} {\  (\bibinfo {year} {2012})},\ \Eprint {http://arxiv.org/abs/1206.5214}
  {arXiv:1206.5214 [hep-lat]}
\bibitem [{\citenamefont {Iwasaki}(1983)}]{Iwasaki:1983ck}%
  \BibitemOpen
  \bibfield  {author} {\bibinfo {author} {\bibfnamefont {Y.}~\bibnamefont
  {Iwasaki}},\ }\href@noop {} {\bibfield  {journal} {\bibinfo  {journal}
  {UTHEP-118}\ } (\bibinfo {year} {1983})}
\bibitem [{\citenamefont {El-Khadra}\ \emph {et~al.}(1997)\citenamefont
  {El-Khadra}, \citenamefont {Kronfeld},\ and\ \citenamefont
  {Mackenzie}}]{ElKhadra:1996mp}%
  \BibitemOpen
  \bibfield  {author} {\bibinfo {author} {\bibfnamefont {A.~X.}\ \bibnamefont
  {El-Khadra}}, \bibinfo {author} {\bibfnamefont {A.~S.}\ \bibnamefont
  {Kronfeld}}, \ and\ \bibinfo {author} {\bibfnamefont {P.~B.}\ \bibnamefont
  {Mackenzie}},\ }\href {\doibase 10.1103/PhysRevD.55.3933} {\bibfield
  {journal} {\bibinfo  {journal} {Phys. Rev.}\ }\textbf {\bibinfo {volume}
  {D55}},\ \bibinfo {pages} {3933} (\bibinfo {year} {1997})},\ \Eprint
  {http://arxiv.org/abs/hep-lat/9604004} {arXiv:hep-lat/9604004}
\bibitem [{\citenamefont {Lin}\ and\ \citenamefont
  {Christ}(2007)}]{Lin:2006ur}%
  \BibitemOpen
  \bibfield  {author} {\bibinfo {author} {\bibfnamefont {H.-W.}\ \bibnamefont
  {Lin}}\ and\ \bibinfo {author} {\bibfnamefont {N.}~\bibnamefont {Christ}},\
  }\href {\doibase 10.1103/PhysRevD.76.074506} {\bibfield  {journal} {\bibinfo
  {journal} {Phys.Rev.}\ }\textbf {\bibinfo {volume} {D76}},\ \bibinfo {pages}
  {074506} (\bibinfo {year} {2007})},\ \Eprint
  {http://arxiv.org/abs/hep-lat/0608005} {arXiv:hep-lat/0608005}
\bibitem [{\citenamefont {Christ}\ \emph {et~al.}(2007)\citenamefont {Christ},
  \citenamefont {Li},\ and\ \citenamefont {Lin}}]{Christ:2006us}%
  \BibitemOpen
  \bibfield  {author} {\bibinfo {author} {\bibfnamefont {N.~H.}\ \bibnamefont
  {Christ}}, \bibinfo {author} {\bibfnamefont {M.}~\bibnamefont {Li}}, \ and\
  \bibinfo {author} {\bibfnamefont {H.-W.}\ \bibnamefont {Lin}},\ }\href
  {\doibase 10.1103/PhysRevD.76.074505} {\bibfield  {journal} {\bibinfo
  {journal} {Phys.Rev.}\ }\textbf {\bibinfo {volume} {D76}},\ \bibinfo {pages}
  {074505} (\bibinfo {year} {2007})},\ \Eprint
  {http://arxiv.org/abs/hep-lat/0608006} {arXiv:hep-lat/0608006}
\bibitem [{\citenamefont {Sheikholeslami}\ and\ \citenamefont
  {Wohlert}(1985)}]{Sheikholeslami:1985ij}%
  \BibitemOpen
  \bibfield  {author} {\bibinfo {author} {\bibfnamefont {B.}~\bibnamefont
  {Sheikholeslami}}\ and\ \bibinfo {author} {\bibfnamefont {R.}~\bibnamefont
  {Wohlert}},\ }\href {\doibase 10.1016/0550-3213(85)90002-1} {\bibfield
  {journal} {\bibinfo  {journal} {Nucl.Phys.}\ }\textbf {\bibinfo {volume}
  {B259}},\ \bibinfo {pages} {572} (\bibinfo {year} {1985})}
\bibitem [{\citenamefont {Aoki}\ \emph {et~al.}(2012)\citenamefont {Aoki},
  \citenamefont {Christ}, \citenamefont {Flynn}, \citenamefont {Izubuchi},
  \citenamefont {Lehner}, \citenamefont {Li}, \citenamefont {Peng},
  \citenamefont {Soni}, \citenamefont {Van~de Water},\ and\ \citenamefont
  {Witzel}}]{Aoki:2012xaa}%
  \BibitemOpen
  \bibfield  {author} {\bibinfo {author} {\bibfnamefont {Y.}~\bibnamefont
  {Aoki}}, \bibinfo {author} {\bibfnamefont {N.~H.}\ \bibnamefont {Christ}},
  \bibinfo {author} {\bibfnamefont {J.~M.}\ \bibnamefont {Flynn}}, \bibinfo
  {author} {\bibfnamefont {T.}~\bibnamefont {Izubuchi}}, \bibinfo {author}
  {\bibfnamefont {C.}~\bibnamefont {Lehner}}, \bibinfo {author} {\bibfnamefont
  {M.}~\bibnamefont {Li}}, \bibinfo {author} {\bibfnamefont {H.}~\bibnamefont
  {Peng}}, \bibinfo {author} {\bibfnamefont {A.}~\bibnamefont {Soni}}, \bibinfo
  {author} {\bibfnamefont {R.~S.}\ \bibnamefont {Van~de Water}}, \ and\
  \bibinfo {author} {\bibfnamefont {O.}~\bibnamefont {Witzel}} (\bibinfo
  {collaboration} {RBC-UKQCD}),\ }\href {\doibase 10.1103/PhysRevD.86.116003}
  {\bibfield  {journal} {\bibinfo  {journal} {Phys. Rev.}\ }\textbf {\bibinfo
  {volume} {D86}},\ \bibinfo {pages} {116003} (\bibinfo {year} {2012})},\
  \Eprint {http://arxiv.org/abs/1206.2554} {arXiv:1206.2554 [hep-lat]}
\bibitem [{\citenamefont {Alford}\ \emph {et~al.}(1996)\citenamefont {Alford},
  \citenamefont {Klassen},\ and\ \citenamefont {Lepage}}]{Alford:1995dm}%
  \BibitemOpen
  \bibfield  {author} {\bibinfo {author} {\bibfnamefont {M.~G.}\ \bibnamefont
  {Alford}}, \bibinfo {author} {\bibfnamefont {T.}~\bibnamefont {Klassen}}, \
  and\ \bibinfo {author} {\bibfnamefont {P.}~\bibnamefont {Lepage}},\ }\href
  {\doibase 10.1016/0920-5632(96)00076-X} {\bibfield  {journal} {\bibinfo
  {journal} {Nucl.Phys.Proc.Suppl.}\ }\textbf {\bibinfo {volume} {47}},\
  \bibinfo {pages} {370} (\bibinfo {year} {1996})},\ \Eprint
  {http://arxiv.org/abs/hep-lat/9509087} {arXiv:hep-lat/9509087 [hep-lat]}
\bibitem [{\citenamefont {Grinstein}\ \emph {et~al.}(1988)\citenamefont
  {Grinstein}, \citenamefont {Springer},\ and\ \citenamefont
  {Wise}}]{Grinstein:1987vj}%
  \BibitemOpen
  \bibfield  {author} {\bibinfo {author} {\bibfnamefont {B.}~\bibnamefont
  {Grinstein}}, \bibinfo {author} {\bibfnamefont {R.~P.}\ \bibnamefont
  {Springer}}, \ and\ \bibinfo {author} {\bibfnamefont {M.~B.}\ \bibnamefont
  {Wise}},\ }\href {\doibase 10.1016/0370-2693(88)90868-4} {\bibfield
  {journal} {\bibinfo  {journal} {Phys. Lett.}\ }\textbf {\bibinfo {volume}
  {B202}},\ \bibinfo {pages} {138} (\bibinfo {year} {1988})}
\bibitem [{\citenamefont {Grinstein}\ \emph {et~al.}(1990)\citenamefont
  {Grinstein}, \citenamefont {Springer},\ and\ \citenamefont
  {Wise}}]{Grinstein:1990tj}%
  \BibitemOpen
  \bibfield  {author} {\bibinfo {author} {\bibfnamefont {B.}~\bibnamefont
  {Grinstein}}, \bibinfo {author} {\bibfnamefont {R.~P.}\ \bibnamefont
  {Springer}}, \ and\ \bibinfo {author} {\bibfnamefont {M.~B.}\ \bibnamefont
  {Wise}},\ }\href {\doibase 10.1016/0550-3213(90)90350-M} {\bibfield
  {journal} {\bibinfo  {journal} {Nucl. Phys.}\ }\textbf {\bibinfo {volume}
  {B339}},\ \bibinfo {pages} {269} (\bibinfo {year} {1990})}
\bibitem [{\citenamefont {Buras}\ \emph {et~al.}(1994)\citenamefont {Buras},
  \citenamefont {Misiak}, \citenamefont {Munz},\ and\ \citenamefont
  {Pokorski}}]{Buras:1993xp}%
  \BibitemOpen
  \bibfield  {author} {\bibinfo {author} {\bibfnamefont {A.~J.}\ \bibnamefont
  {Buras}}, \bibinfo {author} {\bibfnamefont {M.}~\bibnamefont {Misiak}},
  \bibinfo {author} {\bibfnamefont {M.}~\bibnamefont {Munz}}, \ and\ \bibinfo
  {author} {\bibfnamefont {S.}~\bibnamefont {Pokorski}},\ }\href {\doibase
  10.1016/0550-3213(94)90299-2} {\bibfield  {journal} {\bibinfo  {journal}
  {Nucl. Phys.}\ }\textbf {\bibinfo {volume} {B424}},\ \bibinfo {pages} {374}
  (\bibinfo {year} {1994})},\ \Eprint {http://arxiv.org/abs/hep-ph/9311345}
  {arXiv:hep-ph/9311345 [hep-ph]}
\bibitem [{\citenamefont {Ciuchini}\ \emph {et~al.}(1993)\citenamefont
  {Ciuchini}, \citenamefont {Franco}, \citenamefont {Martinelli}, \citenamefont
  {Reina},\ and\ \citenamefont {Silvestrini}}]{Ciuchini:1993ks}%
  \BibitemOpen
  \bibfield  {author} {\bibinfo {author} {\bibfnamefont {M.}~\bibnamefont
  {Ciuchini}}, \bibinfo {author} {\bibfnamefont {E.}~\bibnamefont {Franco}},
  \bibinfo {author} {\bibfnamefont {G.}~\bibnamefont {Martinelli}}, \bibinfo
  {author} {\bibfnamefont {L.}~\bibnamefont {Reina}}, \ and\ \bibinfo {author}
  {\bibfnamefont {L.}~\bibnamefont {Silvestrini}},\ }\href {\doibase
  10.1016/0370-2693(93)90668-8} {\bibfield  {journal} {\bibinfo  {journal}
  {Phys. Lett.}\ }\textbf {\bibinfo {volume} {B316}},\ \bibinfo {pages} {127}
  (\bibinfo {year} {1993})},\ \Eprint {http://arxiv.org/abs/hep-ph/9307364}
  {arXiv:hep-ph/9307364 [hep-ph]}
\bibitem [{\citenamefont {Ciuchini}\ \emph
  {et~al.}(1994{\natexlab{a}})\citenamefont {Ciuchini}, \citenamefont {Franco},
  \citenamefont {Reina},\ and\ \citenamefont {Silvestrini}}]{Ciuchini:1993fk}%
  \BibitemOpen
  \bibfield  {author} {\bibinfo {author} {\bibfnamefont {M.}~\bibnamefont
  {Ciuchini}}, \bibinfo {author} {\bibfnamefont {E.}~\bibnamefont {Franco}},
  \bibinfo {author} {\bibfnamefont {L.}~\bibnamefont {Reina}}, \ and\ \bibinfo
  {author} {\bibfnamefont {L.}~\bibnamefont {Silvestrini}},\ }\href {\doibase
  10.1016/0550-3213(94)90223-2} {\bibfield  {journal} {\bibinfo  {journal}
  {Nucl. Phys.}\ }\textbf {\bibinfo {volume} {B421}},\ \bibinfo {pages} {41}
  (\bibinfo {year} {1994}{\natexlab{a}})},\ \Eprint
  {http://arxiv.org/abs/hep-ph/9311357} {arXiv:hep-ph/9311357 [hep-ph]}
\bibitem [{\citenamefont {Ciuchini}\ \emph
  {et~al.}(1994{\natexlab{b}})\citenamefont {Ciuchini}, \citenamefont {Franco},
  \citenamefont {Martinelli}, \citenamefont {Reina},\ and\ \citenamefont
  {Silvestrini}}]{Ciuchini:1994xa}%
  \BibitemOpen
  \bibfield  {author} {\bibinfo {author} {\bibfnamefont {M.}~\bibnamefont
  {Ciuchini}}, \bibinfo {author} {\bibfnamefont {E.}~\bibnamefont {Franco}},
  \bibinfo {author} {\bibfnamefont {G.}~\bibnamefont {Martinelli}}, \bibinfo
  {author} {\bibfnamefont {L.}~\bibnamefont {Reina}}, \ and\ \bibinfo {author}
  {\bibfnamefont {L.}~\bibnamefont {Silvestrini}},\ }\href {\doibase
  10.1016/0370-2693(94)90602-5} {\bibfield  {journal} {\bibinfo  {journal}
  {Phys. Lett.}\ }\textbf {\bibinfo {volume} {B334}},\ \bibinfo {pages} {137}
  (\bibinfo {year} {1994}{\natexlab{b}})},\ \Eprint
  {http://arxiv.org/abs/hep-ph/9406239} {arXiv:hep-ph/9406239 [hep-ph]}
\bibitem [{\citenamefont {Buras}\ \emph {et~al.}(2002)\citenamefont {Buras},
  \citenamefont {Czarnecki}, \citenamefont {Misiak},\ and\ \citenamefont
  {Urban}}]{Buras:2002tp}%
  \BibitemOpen
  \bibfield  {author} {\bibinfo {author} {\bibfnamefont {A.~J.}\ \bibnamefont
  {Buras}}, \bibinfo {author} {\bibfnamefont {A.}~\bibnamefont {Czarnecki}},
  \bibinfo {author} {\bibfnamefont {M.}~\bibnamefont {Misiak}}, \ and\ \bibinfo
  {author} {\bibfnamefont {J.}~\bibnamefont {Urban}},\ }\href {\doibase
  10.1016/S0550-3213(02)00261-4} {\bibfield  {journal} {\bibinfo  {journal}
  {Nucl. Phys.}\ }\textbf {\bibinfo {volume} {B631}},\ \bibinfo {pages} {219}
  (\bibinfo {year} {2002})},\ \Eprint {http://arxiv.org/abs/hep-ph/0203135}
  {arXiv:hep-ph/0203135 [hep-ph]}
\bibitem [{\citenamefont {Gambino}\ \emph {et~al.}(2003)\citenamefont
  {Gambino}, \citenamefont {Gorbahn},\ and\ \citenamefont
  {Haisch}}]{Gambino:2003zm}%
  \BibitemOpen
  \bibfield  {author} {\bibinfo {author} {\bibfnamefont {P.}~\bibnamefont
  {Gambino}}, \bibinfo {author} {\bibfnamefont {M.}~\bibnamefont {Gorbahn}}, \
  and\ \bibinfo {author} {\bibfnamefont {U.}~\bibnamefont {Haisch}},\ }\href
  {\doibase 10.1016/j.nuclphysb.2003.09.024} {\bibfield  {journal} {\bibinfo
  {journal} {Nucl. Phys.}\ }\textbf {\bibinfo {volume} {B673}},\ \bibinfo
  {pages} {238} (\bibinfo {year} {2003})},\ \Eprint
  {http://arxiv.org/abs/hep-ph/0306079} {arXiv:hep-ph/0306079 [hep-ph]}
\bibitem [{\citenamefont {Altmannshofer}\ \emph {et~al.}(2009)\citenamefont
  {Altmannshofer}, \citenamefont {Ball}, \citenamefont {Bharucha},
  \citenamefont {Buras}, \citenamefont {Straub},\ and\ \citenamefont
  {Wick}}]{Altmannshofer:2008dz}%
  \BibitemOpen
  \bibfield  {author} {\bibinfo {author} {\bibfnamefont {W.}~\bibnamefont
  {Altmannshofer}}, \bibinfo {author} {\bibfnamefont {P.}~\bibnamefont {Ball}},
  \bibinfo {author} {\bibfnamefont {A.}~\bibnamefont {Bharucha}}, \bibinfo
  {author} {\bibfnamefont {A.~J.}\ \bibnamefont {Buras}}, \bibinfo {author}
  {\bibfnamefont {D.~M.}\ \bibnamefont {Straub}}, \ and\ \bibinfo {author}
  {\bibfnamefont {M.}~\bibnamefont {Wick}},\ }\href {\doibase
  10.1088/1126-6708/2009/01/019} {\bibfield  {journal} {\bibinfo  {journal}
  {JHEP}\ }\textbf {\bibinfo {volume} {01}},\ \bibinfo {pages} {019} (\bibinfo
  {year} {2009})},\ \Eprint {http://arxiv.org/abs/0811.1214} {arXiv:0811.1214
  [hep-ph]}
\bibitem [{\citenamefont {Grinstein}\ and\ \citenamefont
  {Pirjol}(2004)}]{Grinstein:2004vb}%
  \BibitemOpen
  \bibfield  {author} {\bibinfo {author} {\bibfnamefont {B.}~\bibnamefont
  {Grinstein}}\ and\ \bibinfo {author} {\bibfnamefont {D.}~\bibnamefont
  {Pirjol}},\ }\href {\doibase 10.1103/PhysRevD.70.114005} {\bibfield
  {journal} {\bibinfo  {journal} {Phys. Rev.}\ }\textbf {\bibinfo {volume}
  {D70}},\ \bibinfo {pages} {114005} (\bibinfo {year} {2004})},\ \Eprint
  {http://arxiv.org/abs/hep-ph/0404250} {arXiv:hep-ph/0404250 [hep-ph]}
\bibitem [{\citenamefont {Beylich}\ \emph {et~al.}(2011)\citenamefont
  {Beylich}, \citenamefont {Buchalla},\ and\ \citenamefont
  {Feldmann}}]{Beylich:2011aq}%
  \BibitemOpen
  \bibfield  {author} {\bibinfo {author} {\bibfnamefont {M.}~\bibnamefont
  {Beylich}}, \bibinfo {author} {\bibfnamefont {G.}~\bibnamefont {Buchalla}}, \
  and\ \bibinfo {author} {\bibfnamefont {T.}~\bibnamefont {Feldmann}},\ }\href
  {\doibase 10.1140/epjc/s10052-011-1635-0} {\bibfield  {journal} {\bibinfo
  {journal} {Eur. Phys. J.}\ }\textbf {\bibinfo {volume} {C71}},\ \bibinfo
  {pages} {1635} (\bibinfo {year} {2011})},\ \Eprint
  {http://arxiv.org/abs/1101.5118} {arXiv:1101.5118 [hep-ph]}
\bibitem [{\citenamefont {Lyon}\ and\ \citenamefont
  {Zwicky}(2014)}]{Lyon:2014hpa}%
  \BibitemOpen
  \bibfield  {author} {\bibinfo {author} {\bibfnamefont {J.}~\bibnamefont
  {Lyon}}\ and\ \bibinfo {author} {\bibfnamefont {R.}~\bibnamefont {Zwicky}},\
  }\href@noop {} {\  (\bibinfo {year} {2014})},\ \Eprint
  {http://arxiv.org/abs/1406.0566} {arXiv:1406.0566 [hep-ph]}
\bibitem [{\citenamefont {Horgan}\ \emph
  {et~al.}(2014{\natexlab{a}})\citenamefont {Horgan}, \citenamefont {Liu},
  \citenamefont {Meinel},\ and\ \citenamefont {Wingate}}]{Horgan:2013pva}%
  \BibitemOpen
  \bibfield  {author} {\bibinfo {author} {\bibfnamefont {R.~R.}\ \bibnamefont
  {Horgan}}, \bibinfo {author} {\bibfnamefont {Z.}~\bibnamefont {Liu}},
  \bibinfo {author} {\bibfnamefont {S.}~\bibnamefont {Meinel}}, \ and\ \bibinfo
  {author} {\bibfnamefont {M.}~\bibnamefont {Wingate}},\ }\href {\doibase
  10.1103/PhysRevLett.112.212003} {\bibfield  {journal} {\bibinfo  {journal}
  {Phys. Rev. Lett.}\ }\textbf {\bibinfo {volume} {112}},\ \bibinfo {pages}
  {212003} (\bibinfo {year} {2014}{\natexlab{a}})},\ \Eprint
  {http://arxiv.org/abs/1310.3887} {arXiv:1310.3887 [hep-ph]}
\bibitem [{\citenamefont {Horgan}\ \emph
  {et~al.}(2014{\natexlab{b}})\citenamefont {Horgan}, \citenamefont {Liu},
  \citenamefont {Meinel},\ and\ \citenamefont {Wingate}}]{Horgan:2013hoa}%
  \BibitemOpen
  \bibfield  {author} {\bibinfo {author} {\bibfnamefont {R.~R.}\ \bibnamefont
  {Horgan}}, \bibinfo {author} {\bibfnamefont {Z.}~\bibnamefont {Liu}},
  \bibinfo {author} {\bibfnamefont {S.}~\bibnamefont {Meinel}}, \ and\ \bibinfo
  {author} {\bibfnamefont {M.}~\bibnamefont {Wingate}},\ }\href {\doibase
  10.1103/PhysRevD.89.094501} {\bibfield  {journal} {\bibinfo  {journal} {Phys.
  Rev.}\ }\textbf {\bibinfo {volume} {D89}},\ \bibinfo {pages} {094501}
  (\bibinfo {year} {2014}{\natexlab{b}})},\ \Eprint
  {http://arxiv.org/abs/1310.3722} {arXiv:1310.3722 [hep-lat]}
\bibitem [{\citenamefont {Horgan}\ \emph {et~al.}(2015)\citenamefont {Horgan},
  \citenamefont {Liu}, \citenamefont {Meinel},\ and\ \citenamefont
  {Wingate}}]{Horgan:2015vla}%
  \BibitemOpen
  \bibfield  {author} {\bibinfo {author} {\bibfnamefont {R.~R.}\ \bibnamefont
  {Horgan}}, \bibinfo {author} {\bibfnamefont {Z.}~\bibnamefont {Liu}},
  \bibinfo {author} {\bibfnamefont {S.}~\bibnamefont {Meinel}}, \ and\ \bibinfo
  {author} {\bibfnamefont {M.}~\bibnamefont {Wingate}},\ }\href@noop {}
  {\bibfield  {journal} {\bibinfo  {journal} {PoS}\ }\textbf {\bibinfo {volume}
  {LATTICE2014}},\ \bibinfo {pages} {372} (\bibinfo {year} {2015})},\ \Eprint
  {http://arxiv.org/abs/1501.00367} {arXiv:1501.00367 [hep-lat]}
\bibitem [{\citenamefont {Bouchard}\ \emph {et~al.}(2013)\citenamefont
  {Bouchard}, \citenamefont {Lepage}, \citenamefont {Monahan}, \citenamefont
  {Na},\ and\ \citenamefont {Shigemitsu}}]{Bouchard:2013mia}%
  \BibitemOpen
  \bibfield  {author} {\bibinfo {author} {\bibfnamefont {C.}~\bibnamefont
  {Bouchard}}, \bibinfo {author} {\bibfnamefont {G.~P.}\ \bibnamefont
  {Lepage}}, \bibinfo {author} {\bibfnamefont {C.}~\bibnamefont {Monahan}},
  \bibinfo {author} {\bibfnamefont {H.}~\bibnamefont {Na}}, \ and\ \bibinfo
  {author} {\bibfnamefont {J.}~\bibnamefont {Shigemitsu}} (\bibinfo
  {collaboration} {HPQCD}),\ }\href {\doibase 10.1103/PhysRevLett.112.149902,
  10.1103/PhysRevLett.111.162002} {\bibfield  {journal} {\bibinfo  {journal}
  {Phys. Rev. Lett.}\ }\textbf {\bibinfo {volume} {111}},\ \bibinfo {pages}
  {162002} (\bibinfo {year} {2013})},\ \bibinfo {note} {[Erratum: Phys. Rev.
  Lett.112,no.14,149902(2014)]},\ \Eprint {http://arxiv.org/abs/1306.0434}
  {arXiv:1306.0434 [hep-ph]}
\bibitem [{\citenamefont {Bailey}\ \emph {et~al.}(2016)\citenamefont {Bailey}
  \emph {et~al.}}]{Bailey:2015dka}%
  \BibitemOpen
  \bibfield  {author} {\bibinfo {author} {\bibfnamefont {J.~A.}\ \bibnamefont
  {Bailey}} \emph {et~al.},\ }\href {\doibase 10.1103/PhysRevD.93.025026}
  {\bibfield  {journal} {\bibinfo  {journal} {Phys. Rev.}\ }\textbf {\bibinfo
  {volume} {D93}},\ \bibinfo {pages} {025026} (\bibinfo {year} {2016})},\
  \Eprint {http://arxiv.org/abs/1509.06235} {arXiv:1509.06235 [hep-lat]}
\bibitem [{\citenamefont {Du}\ \emph {et~al.}(2016)\citenamefont {Du},
  \citenamefont {El-Khadra}, \citenamefont {Gottlieb}, \citenamefont
  {Kronfeld}, \citenamefont {Laiho}, \citenamefont {Lunghi}, \citenamefont
  {Van~de Water},\ and\ \citenamefont {Zhou}}]{Du:2015tda}%
  \BibitemOpen
  \bibfield  {author} {\bibinfo {author} {\bibfnamefont {D.}~\bibnamefont
  {Du}}, \bibinfo {author} {\bibfnamefont {A.~X.}\ \bibnamefont {El-Khadra}},
  \bibinfo {author} {\bibfnamefont {S.}~\bibnamefont {Gottlieb}}, \bibinfo
  {author} {\bibfnamefont {A.~S.}\ \bibnamefont {Kronfeld}}, \bibinfo {author}
  {\bibfnamefont {J.}~\bibnamefont {Laiho}}, \bibinfo {author} {\bibfnamefont
  {E.}~\bibnamefont {Lunghi}}, \bibinfo {author} {\bibfnamefont {R.~S.}\
  \bibnamefont {Van~de Water}}, \ and\ \bibinfo {author} {\bibfnamefont
  {R.}~\bibnamefont {Zhou}},\ }\href {\doibase 10.1103/PhysRevD.93.034005}
  {\bibfield  {journal} {\bibinfo  {journal} {Phys. Rev.}\ }\textbf {\bibinfo
  {volume} {D93}},\ \bibinfo {pages} {034005} (\bibinfo {year} {2016})},\
  \Eprint {http://arxiv.org/abs/1510.02349} {arXiv:1510.02349 [hep-ph]}
\bibitem [{\citenamefont {Hashimoto}\ \emph {et~al.}(1999)\citenamefont
  {Hashimoto}, \citenamefont {El-Khadra}, \citenamefont {Kronfeld},
  \citenamefont {Mackenzie}, \citenamefont {Ryan},\ and\ \citenamefont
  {Simone}}]{Hashimoto:1999yp}%
  \BibitemOpen
  \bibfield  {author} {\bibinfo {author} {\bibfnamefont {S.}~\bibnamefont
  {Hashimoto}}, \bibinfo {author} {\bibfnamefont {A.~X.}\ \bibnamefont
  {El-Khadra}}, \bibinfo {author} {\bibfnamefont {A.~S.}\ \bibnamefont
  {Kronfeld}}, \bibinfo {author} {\bibfnamefont {P.~B.}\ \bibnamefont
  {Mackenzie}}, \bibinfo {author} {\bibfnamefont {S.~M.}\ \bibnamefont {Ryan}},
  \ and\ \bibinfo {author} {\bibfnamefont {J.~N.}\ \bibnamefont {Simone}},\
  }\href {\doibase 10.1103/PhysRevD.61.014502} {\bibfield  {journal} {\bibinfo
  {journal} {Phys. Rev.}\ }\textbf {\bibinfo {volume} {D61}},\ \bibinfo {pages}
  {014502} (\bibinfo {year} {1999})},\ \Eprint
  {http://arxiv.org/abs/hep-ph/9906376} {arXiv:hep-ph/9906376 [hep-ph]}
\bibitem [{\citenamefont {El-Khadra}\ \emph {et~al.}(2001)\citenamefont
  {El-Khadra} \emph {et~al.}}]{ElKhadra:2001rv}%
  \BibitemOpen
  \bibfield  {author} {\bibinfo {author} {\bibfnamefont {A.~X.}\ \bibnamefont
  {El-Khadra}} \emph {et~al.},\ }\href {\doibase 10.1103/PhysRevD.64.014502}
  {\bibfield  {journal} {\bibinfo  {journal} {Phys.Rev.}\ }\textbf {\bibinfo
  {volume} {D64}},\ \bibinfo {pages} {014502} (\bibinfo {year} {2001})},\
  \Eprint {http://arxiv.org/abs/hep-ph/0101023} {arXiv:hep-ph/0101023}
\bibitem [{\citenamefont {Patrignani}\ \emph {et~al.}(2016)\citenamefont
  {Patrignani} \emph {et~al.}}]{Olive:2016xmw}%
  \BibitemOpen
  \bibfield  {author} {\bibinfo {author} {\bibfnamefont {C.}~\bibnamefont
  {Patrignani}} \emph {et~al.} (\bibinfo {collaboration} {Particle Data
  Group}),\ }\href {\doibase 10.1088/1674-1137/40/10/100001} {\bibfield
  {journal} {\bibinfo  {journal} {Chin. Phys.}\ }\textbf {\bibinfo {volume}
  {C40}},\ \bibinfo {pages} {100001} (\bibinfo {year} {2016})}
\bibitem [{\citenamefont {Boyle}\ \emph
  {et~al.}(2016{\natexlab{b}})\citenamefont {Boyle}, \citenamefont
  {Del~Debbio}, \citenamefont {Khamseh}, \citenamefont {Jüttner},
  \citenamefont {Sanfilippo},\ and\ \citenamefont {Tsang}}]{Boyle:2015kyy}%
  \BibitemOpen
  \bibfield  {author} {\bibinfo {author} {\bibfnamefont {P.}~\bibnamefont
  {Boyle}}, \bibinfo {author} {\bibfnamefont {L.}~\bibnamefont {Del~Debbio}},
  \bibinfo {author} {\bibfnamefont {A.}~\bibnamefont {Khamseh}}, \bibinfo
  {author} {\bibfnamefont {A.}~\bibnamefont {Jüttner}}, \bibinfo {author}
  {\bibfnamefont {F.}~\bibnamefont {Sanfilippo}}, \ and\ \bibinfo {author}
  {\bibfnamefont {J.~T.}\ \bibnamefont {Tsang}},\ }\href@noop {} {\bibfield
  {journal} {\bibinfo  {journal} {PoS}\ }\textbf {\bibinfo {volume}
  {LATTICE2015}},\ \bibinfo {pages} {336} (\bibinfo {year}
  {2016}{\natexlab{b}})},\ \Eprint {http://arxiv.org/abs/1511.09328}
  {arXiv:1511.09328 [hep-lat]}
\bibitem [{\citenamefont {Boyle}\ \emph
  {et~al.}(2016{\natexlab{c}})\citenamefont {Boyle}, \citenamefont {Jüttner},
  \citenamefont {Marinkovic}, \citenamefont {Sanfilippo}, \citenamefont
  {Spraggs},\ and\ \citenamefont {Tsang}}]{Boyle:2016imm}%
  \BibitemOpen
  \bibfield  {author} {\bibinfo {author} {\bibfnamefont {P.}~\bibnamefont
  {Boyle}}, \bibinfo {author} {\bibfnamefont {A.}~\bibnamefont {Jüttner}},
  \bibinfo {author} {\bibfnamefont {M.~K.}\ \bibnamefont {Marinkovic}},
  \bibinfo {author} {\bibfnamefont {F.}~\bibnamefont {Sanfilippo}}, \bibinfo
  {author} {\bibfnamefont {M.}~\bibnamefont {Spraggs}}, \ and\ \bibinfo
  {author} {\bibfnamefont {J.~T.}\ \bibnamefont {Tsang}},\ }\href {\doibase
  10.1007/JHEP04(2016)037} {\bibfield  {journal} {\bibinfo  {journal} {JHEP}\
  }\textbf {\bibinfo {volume} {04}},\ \bibinfo {pages} {037} (\bibinfo {year}
  {2016}{\natexlab{c}})},\ \Eprint {http://arxiv.org/abs/1602.04118}
  {arXiv:1602.04118 [hep-lat]}
\bibitem [{\citenamefont {Boyle}\ \emph
  {et~al.}(2016{\natexlab{d}})\citenamefont {Boyle}, \citenamefont
  {Del~Debbio}, \citenamefont {Jüttner}, \citenamefont {Khamseh},
  \citenamefont {Sanfilippo}, \citenamefont {Tsang},\ and\ \citenamefont
  {Witzel}}]{Boyle:2016lzk}%
  \BibitemOpen
  \bibfield  {author} {\bibinfo {author} {\bibfnamefont {P.}~\bibnamefont
  {Boyle}}, \bibinfo {author} {\bibfnamefont {L.}~\bibnamefont {Del~Debbio}},
  \bibinfo {author} {\bibfnamefont {A.}~\bibnamefont {Jüttner}}, \bibinfo
  {author} {\bibfnamefont {A.}~\bibnamefont {Khamseh}}, \bibinfo {author}
  {\bibfnamefont {F.}~\bibnamefont {Sanfilippo}}, \bibinfo {author}
  {\bibfnamefont {J.~T.}\ \bibnamefont {Tsang}}, \ and\ \bibinfo {author}
  {\bibfnamefont {O.}~\bibnamefont {Witzel}},\ }\href@noop {} {\  (\bibinfo
  {year} {2016}{\natexlab{d}})},\ \Eprint {http://arxiv.org/abs/1611.06804}
  {arXiv:1611.06804 [hep-lat]}
\bibitem [{\citenamefont {Cho}\ \emph {et~al.}(2015)\citenamefont {Cho},
  \citenamefont {Hashimoto}, \citenamefont {Jüttner}, \citenamefont {Kaneko},
  \citenamefont {Marinkovic}, \citenamefont {Noaki},\ and\ \citenamefont
  {Tsang}}]{Cho:2015ffa}%
  \BibitemOpen
  \bibfield  {author} {\bibinfo {author} {\bibfnamefont {Y.-G.}\ \bibnamefont
  {Cho}}, \bibinfo {author} {\bibfnamefont {S.}~\bibnamefont {Hashimoto}},
  \bibinfo {author} {\bibfnamefont {A.}~\bibnamefont {Jüttner}}, \bibinfo
  {author} {\bibfnamefont {T.}~\bibnamefont {Kaneko}}, \bibinfo {author}
  {\bibfnamefont {M.}~\bibnamefont {Marinkovic}}, \bibinfo {author}
  {\bibfnamefont {J.-I.}\ \bibnamefont {Noaki}}, \ and\ \bibinfo {author}
  {\bibfnamefont {J.~T.}\ \bibnamefont {Tsang}},\ }\href {\doibase
  10.1007/JHEP05(2015)072} {\bibfield  {journal} {\bibinfo  {journal} {JHEP}\
  }\textbf {\bibinfo {volume} {05}},\ \bibinfo {pages} {072} (\bibinfo {year}
  {2015})},\ \Eprint {http://arxiv.org/abs/1504.01630} {arXiv:1504.01630
  [hep-lat]}
\bibitem [{\citenamefont {Boyle}\ \emph
  {et~al.}(2016{\natexlab{e}})\citenamefont {Boyle}, \citenamefont
  {Del~Debbio},\ and\ \citenamefont {Khamseh}}]{Boyle:2016mwo}%
  \BibitemOpen
  \bibfield  {author} {\bibinfo {author} {\bibfnamefont {P.}~\bibnamefont
  {Boyle}}, \bibinfo {author} {\bibfnamefont {L.}~\bibnamefont {Del~Debbio}}, \
  and\ \bibinfo {author} {\bibfnamefont {A.}~\bibnamefont {Khamseh}},\
  }\href@noop {} {\  (\bibinfo {year} {2016}{\natexlab{e}})},\ \Eprint
  {http://arxiv.org/abs/1608.07982} {arXiv:1608.07982 [hep-lat]}
\bibitem [{\citenamefont {Boyle}\ \emph
  {et~al.}(2016{\natexlab{f}})\citenamefont {Boyle}, \citenamefont
  {Del~Debbio},\ and\ \citenamefont {Khamseh}}]{Boyle:2016wis}%
  \BibitemOpen
  \bibfield  {author} {\bibinfo {author} {\bibfnamefont {P.}~\bibnamefont
  {Boyle}}, \bibinfo {author} {\bibfnamefont {L.}~\bibnamefont {Del~Debbio}}, \
  and\ \bibinfo {author} {\bibfnamefont {A.}~\bibnamefont {Khamseh}},\
  }\href@noop {} {\  (\bibinfo {year} {2016}{\natexlab{f}})},\ \Eprint
  {http://arxiv.org/abs/1611.06908} {arXiv:1611.06908 [hep-lat]}
\end{thebibliography}%

%

\end{document}